\documentclass[graybox, secnum]{svmult}

\usepackage{mathptmx}       
\usepackage{helvet}         
\usepackage{courier}        
\usepackage{type1cm}        
\usepackage{makeidx}         
\usepackage{graphicx}        
\usepackage{multicol}        
\usepackage[bottom]{footmisc}
\usepackage{hyperref}        
\usepackage{soul}            
\usepackage[percent]{overpic}
\hypersetup{colorlinks=true,urlcolor=blue}
\usepackage[square,numbers]{natbib}
\usepackage{amsmath,amssymb}
\usepackage[T1]{fontenc}

\newcommand\msun{{\rm M_{\odot}}}

\def\go{
\mathrel{\raise.3ex\hbox{$>$}\mkern-14mu\lower0.6ex\hbox{$\sim$}}
}
\def\lo{
\mathrel{\raise.3ex\hbox{$<$}\mkern-14mu\lower0.6ex\hbox{$\sim$}}
}

\newcommand{\chandra}{\textit{Chandra}}
\newcommand{\rxte}{\textit{RXTE}}

\newcommand{\rosat}{\textit{ROSAT}}

\newcommand{\suzaku}{\textit{Suzaku}}
\newcommand{\xmm}{\textit{XMM-Newton}}

\newcommand{\swift}{\textit{Swift}}

\newcommand{\nustar}{\textit{NuSTAR}}

\newcommand{\ms}{\ensuremath{M_{\odot}}}

\newcommand{\fluxcgs}{\ensuremath{\mathrm{erg}\,\mathrm{cm}^{-2}\,\mathrm{s}^{-1}\,}}

\newcommand{\rmag}{$R_{\rm m}$}
\newcommand{\pspin}{$P_{\rm spin}$}
\newcommand{\pbeat}{$P_{\rm beat}$}
\newcommand{\porb}{$P_{\rm orb}$}
\newcommand{\kts}{$kT_{\rm shock}$}
\newcommand{\mwd}{$M_{\rm WD}$}
\newcommand{\rwd}{$R_{\rm WD}$}
\newcommand{\avmwd}{$\bar{M}_{\rm WD}$}

\makeindex             

\begin{document}
\title*{X-ray emission mechanisms in accreting white dwarfs}
\author{K.L. Page \thanks{corresponding author} and A.W. Shaw}
\institute{K.L. Page \at School of Physics \& Astronomy, University of Leicester, LE1 7RH, UK, \email{klp5@leicester.ac.uk}
\and A.W. Shaw \at Department of Physics, University of Nevada, Reno, NV 89557, USA \email{aarrans@unr.edu}}
%
%
\maketitle
\abstract{In this chapter we consider the processes which can lead to X-ray emission from different types of cataclysmic variable stars (CVs). CVs are semi-detached, binary star systems where material is transferred from the donor star (also known as the companion or secondary star) onto the white dwarf primary. CVs are divided into several sub-classes based on the observed phenomenology in the optical and X-ray bands, which, in turn, is largely defined by the magnetic field strength of the accretor. In non-magnetic systems, a variety of observed behaviours are identified, depending on the accretion rate: novae, dwarf novae, nova-like variables, symbiotic binaries and supersoft sources are all examples of non-magnetic CVs. In magnetic systems (polars and intermediate polars, or AM Her and DQ Her systems, respectively), the accretion flow is channelled to polar regions, and the observational appearance is different. X-rays are typically produced through hot or energetic processes, and in CVs they are formed via shocks (within a boundary layer or accretion column,  or through interactions either internal to the nova ejecta, or between the ejecta and a stellar wind) or from hydrogen burning (either steady fusion, or a thermonuclear runaway). All of these different types of accreting white dwarfs are discussed here, considering both spectral and temporal variability in the different populations.}

\section{Keywords} 
Cataclysmic variable stars -- X-ray binary stars -- Novae -- Dwarf novae -- Nova-like variables -- Persistent super-soft sources -- Symbiotic binary stars -- Polars -- Intermediate Polars -- AM Canum Venaticorum stars

\section{Introduction}

As discussed in the chapter by Webb, there are many different types of accreting white dwarfs (WDs); this classification is historically based predominantly on phenomenology observed in the optical band. Here we consider the mechanisms through which these systems may produce X-rays. We emphasize that the physical processes responsible for both the X-ray and optical emission are tightly interconnected, so it is important to understand both of them. As such, use of the commonly-used CV classification is still justified.
In general, the energetic X-rays are formed on, or very close to, the WD itself  (determined via observations of eclipsing systems; e.g. \cite{mukai97}), whereas the optical emission comes from further out in the binary system, and more than one mechanism for forming X-rays may be active in any given type of cataclysmic variable (CV) system. Both optically-thin and optically-thick X-ray emission is possible, depending on the processes occurring. Continuum and line emission (or absorption) is measured, and this can be frequently absorbed -- totally or partially -- by the surrounding environment.

Accreting WDs can be split into non-magnetic and magnetic sources, with the cut being a field strength of around 10$^{6}$~G. In this chapter we will first cover the non-magnetic population, before progressing to the sources with significant magnetic fields. There is, of course, no hard start or end point for the X-ray bandpass within the electromagnetic
spectrum. Here we consider energies from around 0.3 keV up to $\sim$~150~keV, which can be observed by current missions such as {\it Swift} (XRT\footnote{X-Ray Telescope} and BAT\footnote{Burst Alert Telescope}), {\it MAXI}\footnote{Monitor of All-sky X-ray Image}, \nustar\footnote{Nuclear Spectroscopic Telescope Array} and {\it NICER}\footnote{Neutron star Interior Composition Explorer}, for example. 

In addition to the information in this chapter, \cite{Mukai-2017} provides an excellent review on the subject of X-rays from accreting WDs.

 \section{Novae}
\label{nova}

As accreting WDs go, novae are the most explosive type. Mass transfer from the secondary star to the WD leads to the formation of a layer of hydrogen on the WD surface. Eventually, when enough material has been transferred, nuclear burning will ignite at the base of this envelope; the pressure will then increase, until it reaches a sufficient level to trigger a thermonuclear runaway  (TNR; see review articles in \cite{bodeevans08}). Following this TNR explosion, material is flung outward, obscuring the WD surface. To date, this point is usually the first sign of a nova eruption, when a new optical source is detected: the optical peak occurs at the point of maximum expansion of the photosphere\footnote{While most novae are first discovered as new optical sources, V959~Mon was initially detected in $\gamma$-rays by the {\it Fermi}-LAT, when too close to the Sun for ground-based telescopes to observe; an optical counterpart was later discovered, and the two detections identified as being one and the same source \cite{cheung12b}.}. However, models of nova outbursts predict that there should be a brief (0.5+ day), soft X-ray flash shortly after hydrogen ignition, but before the optical discovery, known as the `fireball phase' \cite[e.g.][]{starrfield90}. While several searches have previously been attempted \cite[e.g.][]{kato16, morii16}, it is only recently that such an X-ray flash has been detected -- by {\em eROSITA}\footnote{extended Roentgen Survey with an Imaging Telescope Array} during its second all-sky survey \citep{konig22}, for the nova YZ~Ret (Nova Ret 2020). In this case, a very soft flash of X-rays was detected 11~hr before its optical brightening, lasting no more than 8~hr. The softness of the spectrum is likely the reason why {\em MAXI} failed to find any similar flashes \cite{morii16}.
While there are many ground-based telescopes -- both large facilities and the far-reaching community of amateur astronomers -- constantly scanning for new optical transients, we are currently lacking in all-sky X-ray monitors to provide the same level of prompt transient location. In the future, missions such as {\it Einstein Probe} (currently planned for launch at the end of 2022), will observe the entire sky at X-ray energies with a high cadence, hopefully allowing us to identify more such X-ray flashes. 

As the nova ejecta expand, they become optically thin, usually allowing the surface nuclear burning to become visible\footnote{If the nuclear burning only lasts for a very short interval, it may switch off before the ejecta thin out enough to allow the soft X-rays to become visible. In the case of V745~Sco \cite{page15}, it is postulated that the active nuclear burning had ended before the ejecta had fully cleared, meaning that only the cooling emission was later detected, so placing it close to this `invisibility zone' where the SSS emission would be entirely missed.}. This optically-thick radiation peaks in the soft X-ray band, and is thus termed the `super-soft source' (SSS) phase \cite{krautter08}. This soft emission continues for as long as there is sufficient hydrogen for nuclear burning, and can sometimes be extended longer than expected if accretion resumes early on \cite[e.g.][]{aydi18}.
A cooling and fading soft spectral component may still be seen for a while after the burning has come to an end; when the nuclear reactions can no longer be sustained, the nova returns to quiescence. 
Once accretion resumes, hydrogen again builds up on the WD surface; thus the nova cycle begins anew. Classical novae do not tend to be detectable X-ray emitters during quiescence. Recurrent novae\footnote{Classical novae are those which have only been seen to erupt once; a small number of systems have been detected in outburst multiple times, and these are known as recurrent novae. All novae are expected to erupt more than once, but over timescales of typically thousands of years.}, on the other hand, have higher accretion rates \cite{brad10}, and can often be detected in X-rays between eruptions \cite[e.g.][]{orio93}.

\subsection{X-ray light curves of novae}
\label{nova:lc}

\begin{figure}[t]
\begin{center}

  \includegraphics[clip, angle=-90, width=10cm]{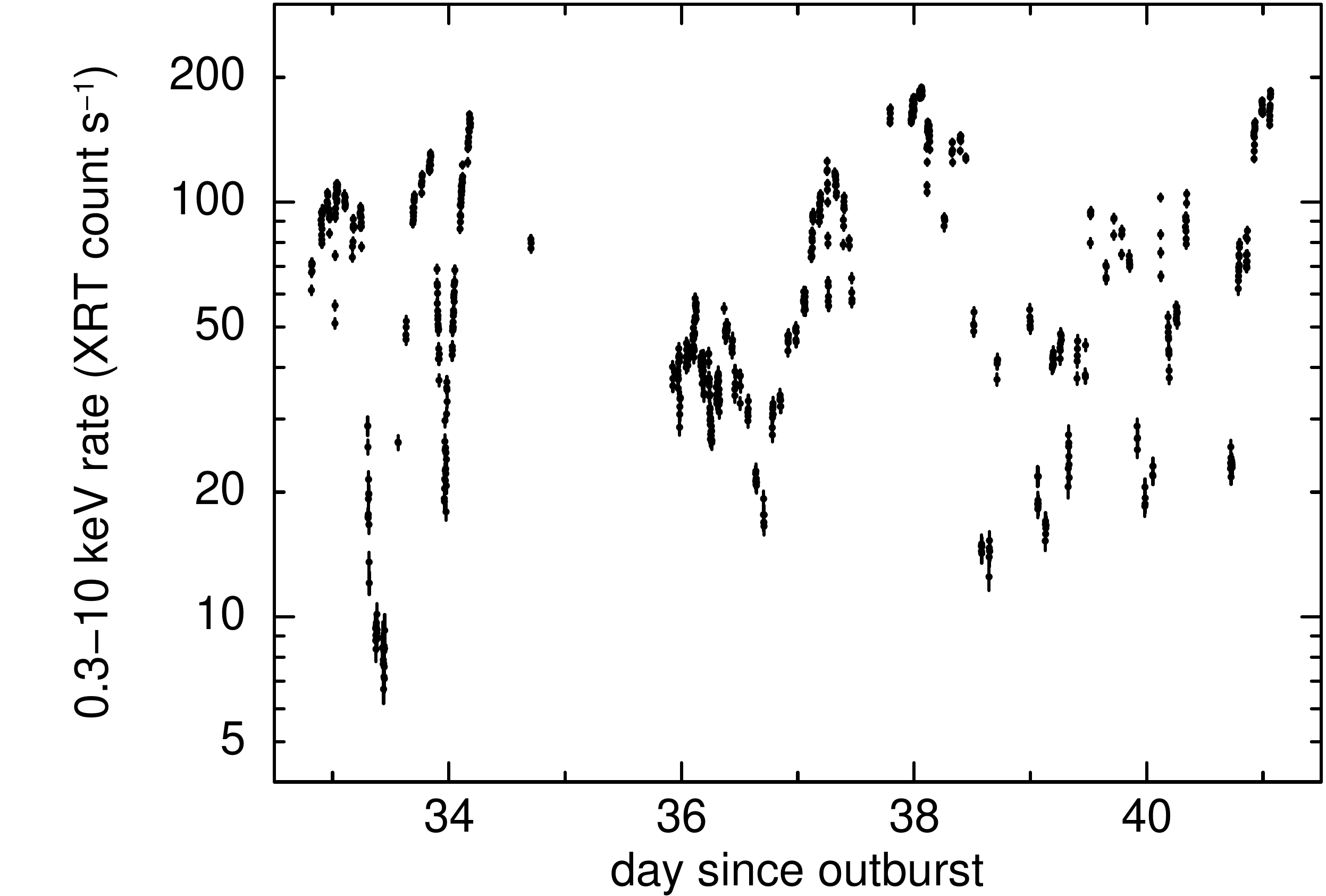}
  \includegraphics[clip, angle=-90, width=10cm]{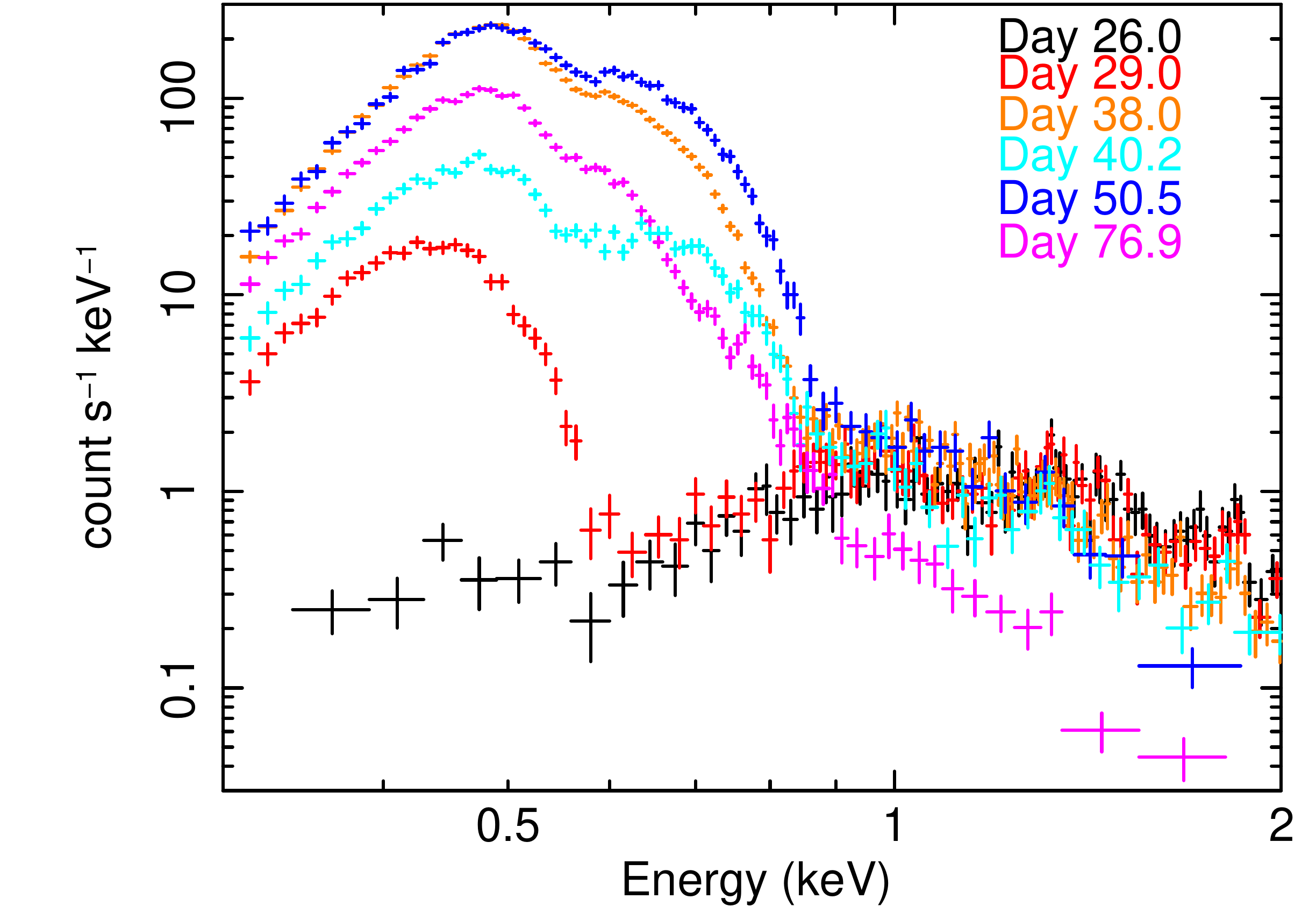}
\caption{Top: The beginning of the super-soft X-ray emission frequently shows large scale flux variability. This was first identified in the {\it Swift} observations of RS Oph following its 2006 eruption, shown here. Bottom: A sample of SSS X-ray spectra for RS Oph, showing a variety of temperatures and brightnesses. The legend gives the days after the 2006 eruption when the spectra were obtained.}
\label{var}
\end{center}
\end{figure}

\begin{figure}
\begin{center}
\includegraphics[clip, angle=-90, width=10cm]{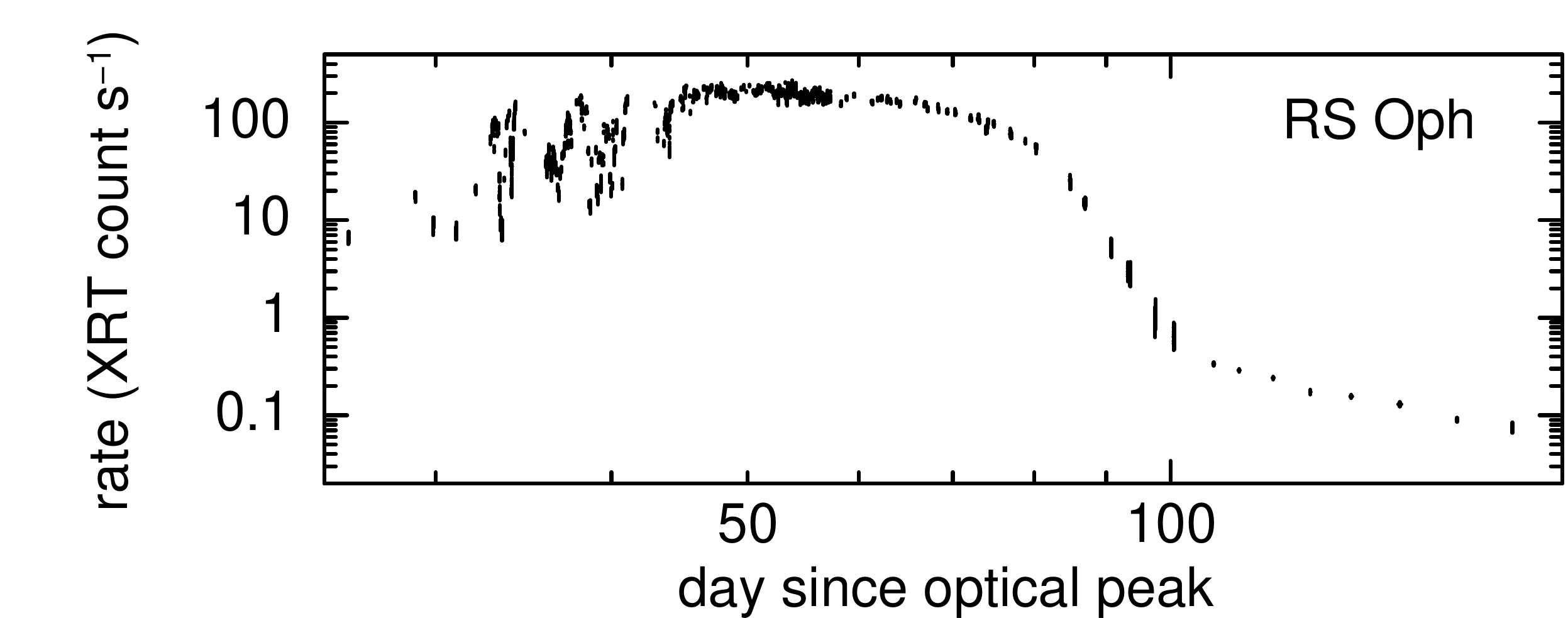}
\includegraphics[clip, angle=-90, width=10cm]{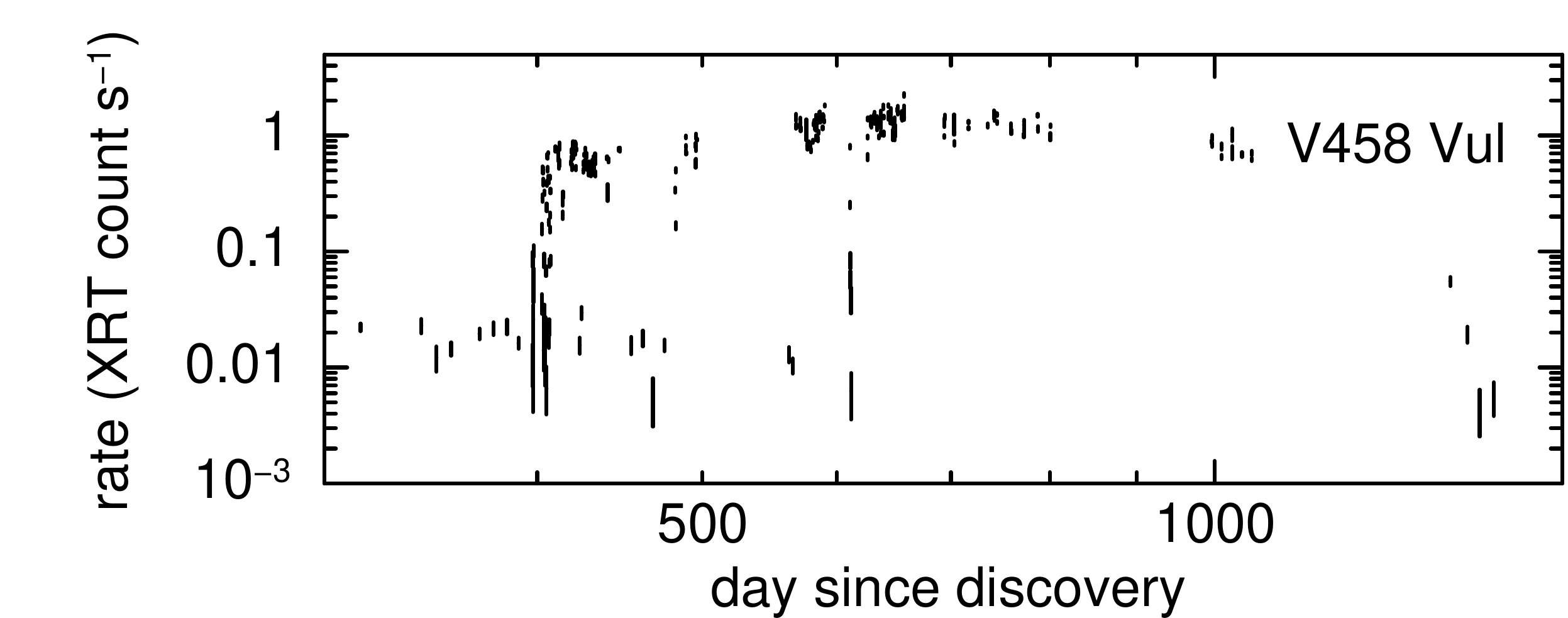}
\includegraphics[clip, angle=-90, width=10cm]{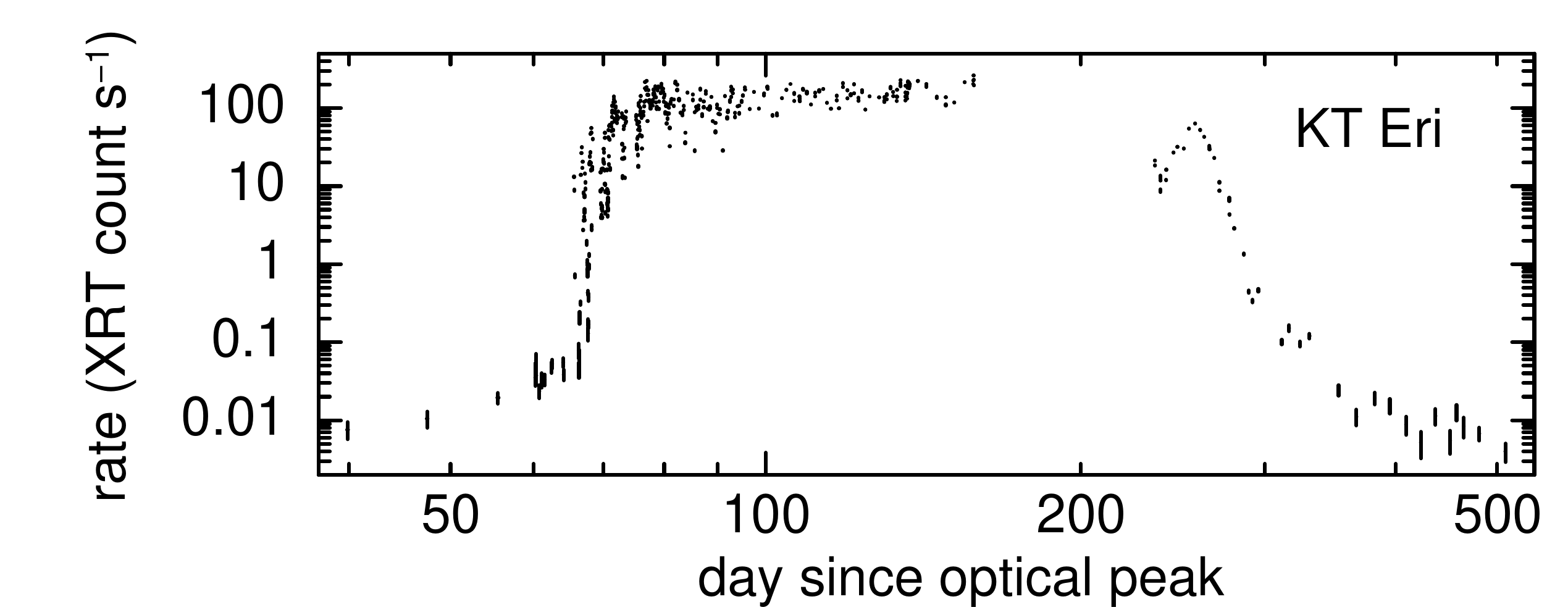}
\includegraphics[clip, angle=-90, width=10cm]{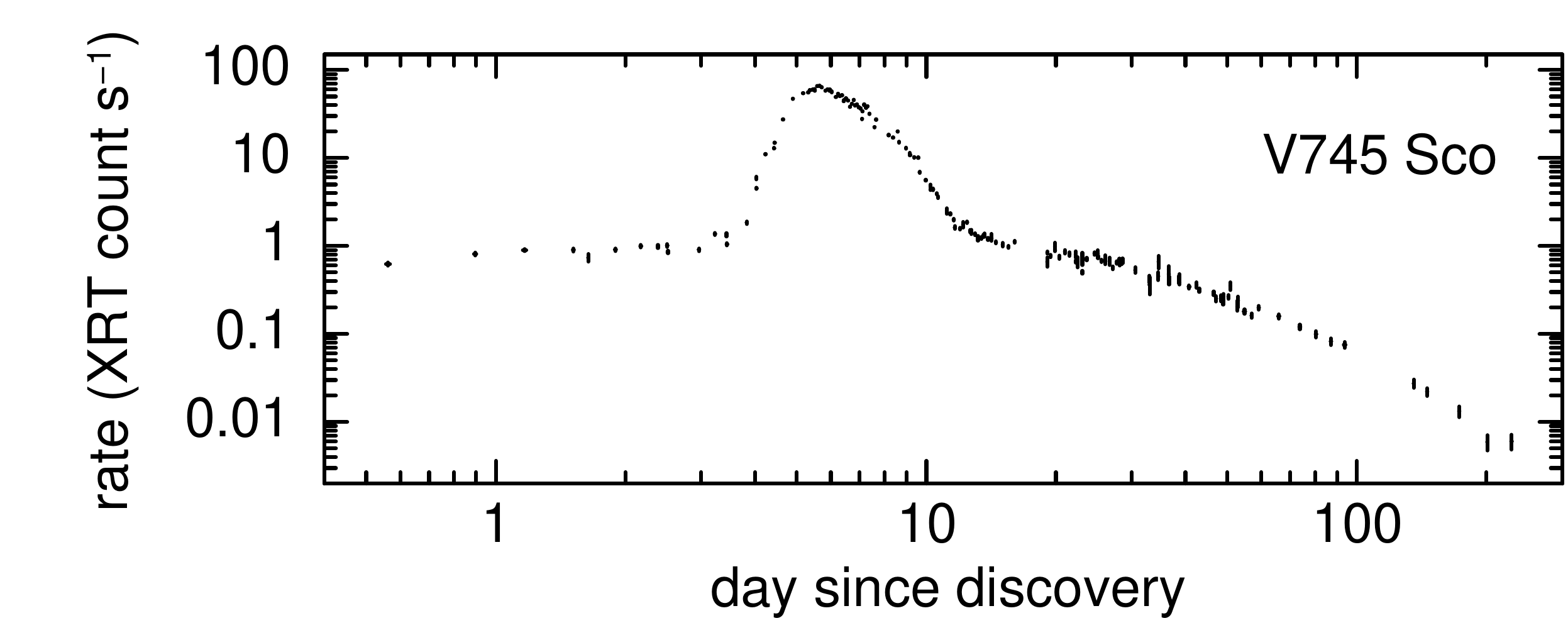}
\caption{A selection of well-monitored {\em Swift}-XRT nova light curves, showing the different onsets of the SSS phase, some smooth, some chaotic, and subsequent evolution. The complete datasets are not always shown, in order to highlight the more interesting intervals.}
\label{nova-lc}
\end{center}
\end{figure}

The onset of the SSS phase is often found to be chaotic when monitored at a sufficiently high frequency (at least several times a day); this is demonstrated in the top panel of Fig.~\ref{var}. Such variability is not always the case, though, and \cite{page15} discusses a clear counterexample; see also the bottom panel of Fig.~\ref{nova-lc}.
This high-amplitude flux variability was first identified in the {\it Swift} observations of the 2006 eruption of RS~Oph \cite{julo11}, but subsequently found in many other novae (see \cite{page20} for a review). While the exact details of the mechanism leading to this variability are uncertain, it appears to be at least partly due to variable visibility of the WD \cite[e.g.][]{julo11, page14}. Following the nova eruption, the material ejected may well be clumpy. Should these clumps pass through the observer's line of sight, soft X-rays (which dominate the spectrum of the source at this time) will be blocked, and the measured count rate drop. Spectral fits also show variations in the photospheric temperature; measurements of $<$~30 up to $>$100~eV are found when fitting {\it Swift} X-ray spectra  \cite[e.g.][]{page14}, with the lowest temperatures leading to some -- possibly most -- of the SSS emission occurring outside the XRT bandpass, and thus a decreased observed count rate over 0.3--10~keV.

Fig.~\ref{nova-lc} presents a selection of X-ray light curves of novae well-monitored by {\em Swift}, demonstrating the range of variability detected; see \cite{page20} for a large sample.

\begin{figure}
\begin{center}

  \includegraphics[clip, angle=-90, width=5.5cm]{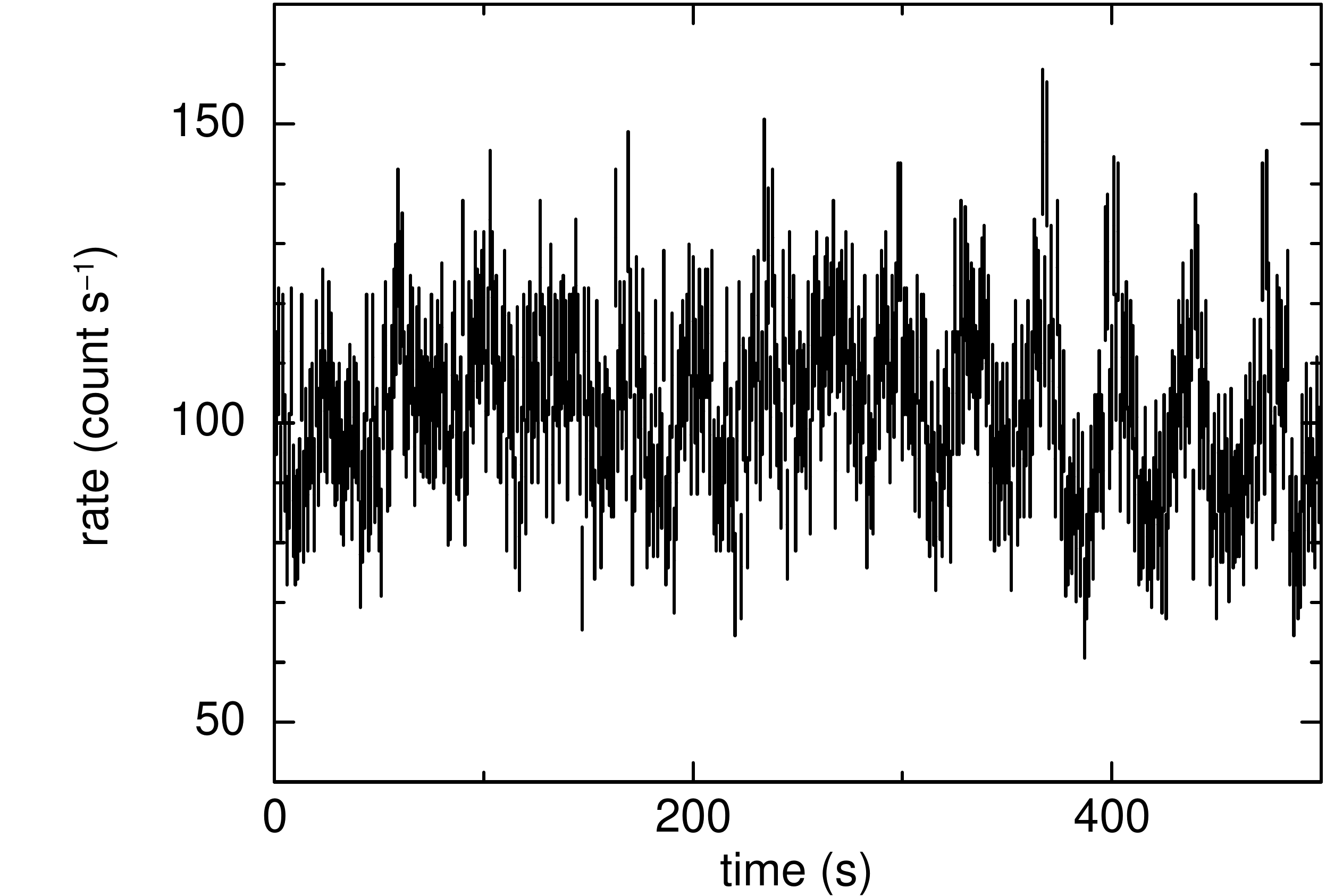}
  \includegraphics[clip, angle=-90, width=5.5cm]{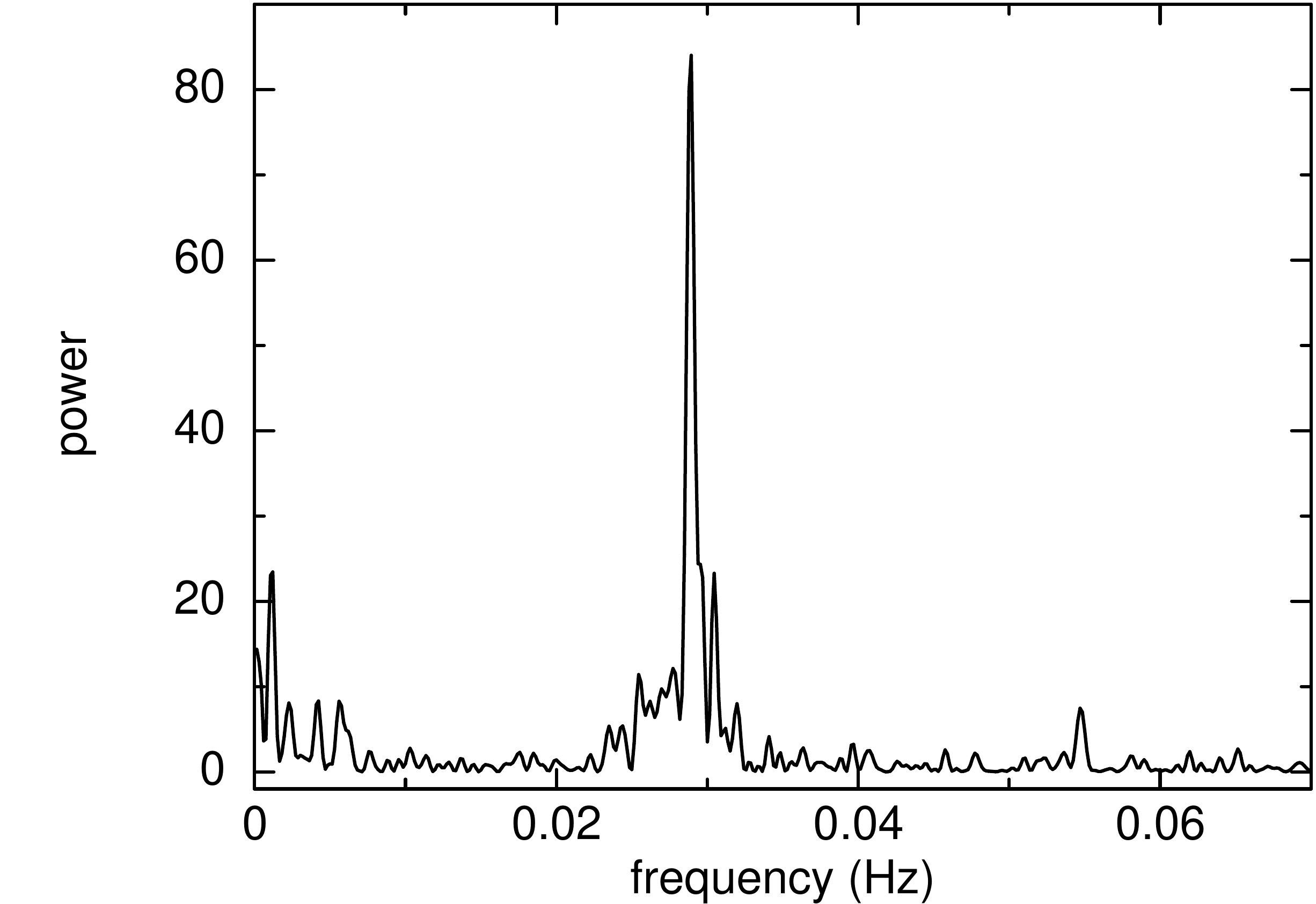}
\caption{Left: An interval during which the $\sim$~35~s X-ray QPO was particularly strong in the 2006 eruption soft (0.3--1 keV) light curve of RS Oph. Right: Periodogram corresponding to this short interval of data.}
\label{qpo}
\end{center}
\end{figure}

In addition to this impressive high-amplitude variability in X-ray brightness, the RS~Oph data showed a strong, smaller-scale quasi-periodic oscillation (QPO) of around 35~s during much of the bright SSS phase \cite{julo11}; Fig.~\ref{qpo} highlights one of the intervals of interest. Following this initial discovery, transient QPOs of up to $\sim$~100~s were subsequently identified in other X-ray bright, SSS phases of novae, as well \cite[e.g.][]{ness15}. It is not yet certain what drives these oscillations, with explanations ranging from rotation to non-radial g-mode pulsations (that is, low-frequency non-radial, gravity mode -- or buoyancy -- pulsations, driven by temperature-sensitive nuclear burning; \cite{kawaler88}); see discussion in \cite{ness15}. All the suggestions present certain problems, however. If the modulation is related to rotation, then some form of differentiation of the emission is required, rather than the nuclear burning being constant over the entire WD surface, possible examples being occultation of a magnetic hotspot, or variable absorption. However, the oscillations sometimes appear multi-periodic, and the fractional amplitude varies, which cannot be explained by a {\it simple} WD rotation model \citep{beardmore08}.
While the g-mode pulsations were suggested as an explanation by \cite{julo11}, more recent work by \cite{wolf18} predicts that pulsations would only be stable under $\sim$~10s, shorter than the QPOs which have been measured in novae.

\subsection{X-ray spectra of novae}

\begin{figure}
\begin{center}

  \includegraphics[clip, width=12cm]{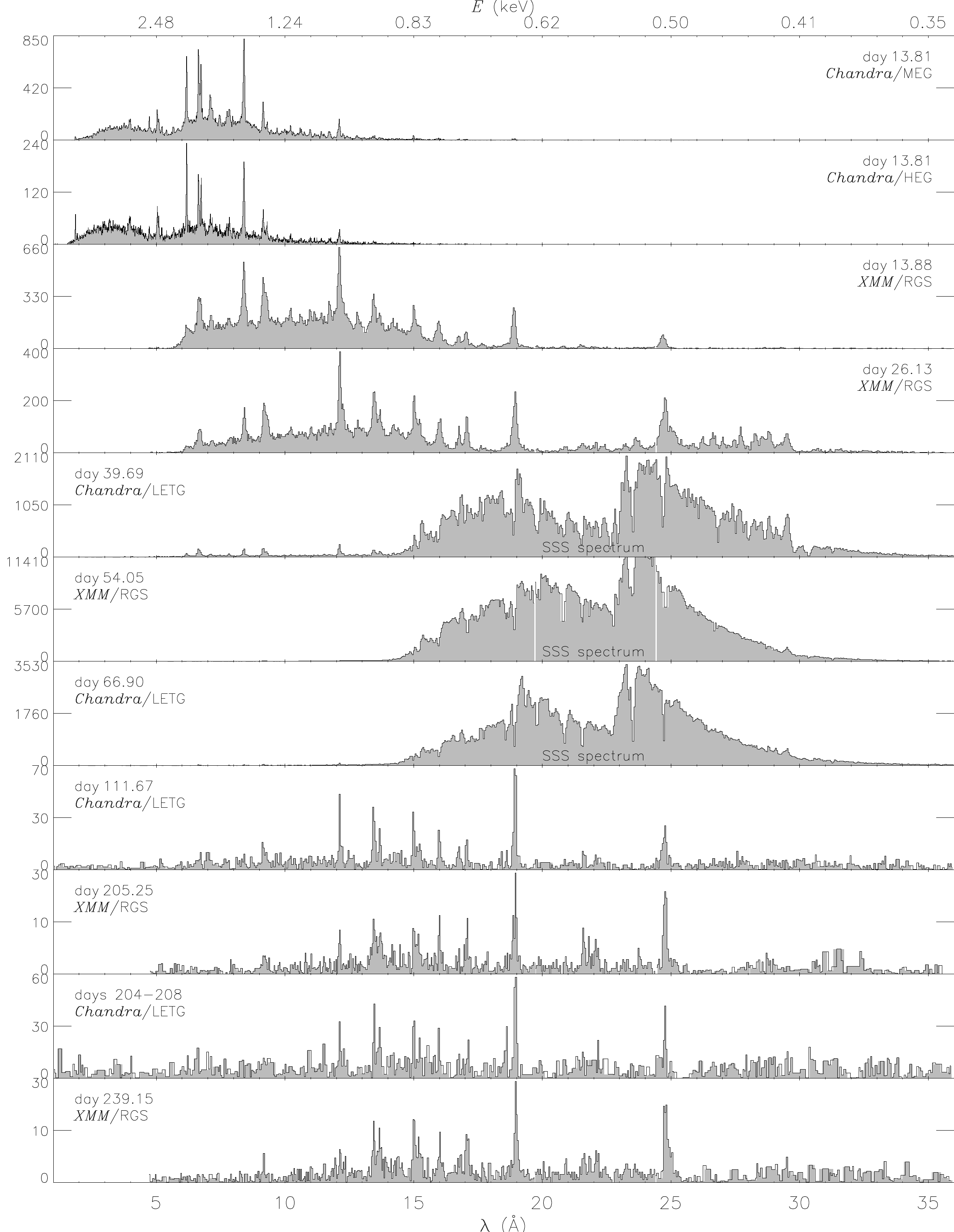}
  \caption{An sample of X-ray grating spectra obtained with \chandra\ and \xmm\ following the 2006 eruption of RS~Oph, showing a variety of emission and absorption features -- clearly more complex than a simple BB continuum as low-resolution spectra can suggest. Reproduced from \cite{ness09}.}
\label{grating}
\end{center}
\end{figure}

The SSS emission appears, to first order, similar to a blackbody (BB; lower panel of Fig.~\ref{var}), possibly with superimposed absorption edges, particularly when using low-resolution spectra from an instrument such as the XRT onboard {\it Swift}. Grating data, from the Reflection Grating Spectrometer (RGS) on \xmm\ or the \chandra\ Low/High Energy Transmission Gratings (LETG, HETG), show a much more complicated situation with multiple emission and/or absorption lines superimposed on the continuum (Fig.~\ref{grating}). It is clear that a BB is a vast over-simplification of the actual emission, which must be more akin to a stellar atmosphere. Indeed, using a BB approximation can potentially underestimate the temperature and overestimate the luminosity \cite{krautter96}, although, as shown by \cite{julo11}, this is not always the case. However, the currently (2022) available idealised stellar atmosphere models tend to provide worse fits from a statistical standpoint than a simple BB, and the complex results from high-resolution X-ray spectra indicate further development of atmosphere models is needed before conclusive, physically-motivated modelling of nova X-ray spectra can be performed \cite{ness20}.

While the soft, optically-thick emission discussed above dominates the spectra during the SSS phase, harder X-rays arising from shock interactions can also be observed before, during and after this interval \cite[e.g.,][]{brecher77}.
As reviewed by \cite{chomiuk21}, shocked X-ray emission in classical novae can be caused by collisions between outflows moving at different velocities.

In the case of recurrent novae, the evolved companions will be losing material in the form of a wind. The material ejected from the WD during the nova eruption should collide with this wind, leading to shocked emission. \cite{bode06} demonstrated that the data collected for the 1985 and 2006 eruptions of RS~Oph were consistent with shocks propagating through the wind from the red giant (RG) companion. This shocked emission can typically be parameterised in X-ray spectra by a small number of discrete temperature optically-thin components, though this is a simplification of a much more complex situation \cite[e.g.][]{vaytet11}.

In some systems, the signatures of shocks in novae are seen to extend beyond the X-ray range considered in this chapter, and up to GeV energies. This is discussed in the next section.

We also note that novae can occur in magnetic systems (polars and intermediate polars -- IPs; see Section~\ref{sec:mag} for a detailed discussion on X-rays from these sources), which may lead to differences in the observed X-ray emission, with V407 Lup (Nova Lupi 2016) being a well-monitored example \cite{aydi18}. There is an indication in this nova that harder X-rays may `turn on' at late times, whereas shock emission would be expected to be present from the start. This, together with the detection of two distinct periodicities in the optical, UV and X-ray data, was taken as evidence of the likely magnetic nature of the system.

\subsection{Higher energies}

Since the 2008 launch of the {\it Fermi Gamma-ray Space Telescope}, with its Large Area Telescope (LAT) covering energies from $\sim$~20~MeV to $>$300~GeV, more and more novae have been detected at GeV energies. \cite{gordon20} consider {\it Fermi}-LAT-detected novae which were also observed by {\it Swift}-XRT, finding that, somewhat surprisingly, the X-rays expected from shocks are not usually visible concurrently with the GeV emission, but rather appear after the $\gamma$-rays have faded away. V407~Cyg, with its RG companion, was the exception (though V745~Sco and V1535~Sco -- each with giant companions -- were both also clearly detected in X-rays at early times, but only had marginal LAT detections); all other novae in the sample were systems with dwarf secondary stars. The authors suggest a scenario whereby the GeV emission in recurrent systems with giant stars is related to external shocks between the WD ejecta and the dense wind from the companion, while the novae with Main Sequence companions produce $\gamma$-rays via internal shocks between multiple ejection events \cite{chomiuk14}; these latter shocks are embedded within the high-density ejecta, and thus initially hidden from view. Alternatively, the X-rays may be suppressed early on by corrugated shock fronts \cite{metzger15, steinberg18}. 
Harder X-rays ($>$~10 keV) are less likely to be absorbed, however, and observations by \nustar\ have, indeed, identified X-ray emission concurrent with GeV $\gamma$-rays \citep[e.g. ][]{tommy19}, though fainter than might be expected from the corresponding $\gamma$-ray luminosities \cite{gordon20}.

\section{Dwarf Novae}

Despite the similar names, dwarf novae (DNe) and novae are very different beasts. While a nova eruption is related to processes on the WD itself (Section~\ref{nova}), the preferred explanation for DNe is the so-called Disc Instability Model (DIM), where the outburst is caused by changes in the state of the accretion disc. In short, material transferred from the secondary star builds up in the disc more rapidly than it can be accreted onto the WD; this continues until the density and temperature reach critical levels corresponding to partial hydrogen ionisation in the inner disc region. This leads to an increase in the viscosity, and hence a disc instability, which causes much of the excess material to be rapidly accreted, releasing a burst of gravitational potential energy.
There is also an alternative proposed model, wherein the outburst is caused by a sudden increase in the mass transfer rate from the companion star to the disc (that is, an instability in the secondary star itself leads to the DN outburst); however, a physical mechanism leading to this stellar instability is not altogether obvious. See chapter by Davis for more details on DN optical outbursts and the DIM. It is interesting to note that the DIM has also been applied to explain some parts of X-ray pulsar phenomenology \cite{tsygankov17} with the only difference being that the inner disc radius is defined by the magnetosphere of the neutron star rather than the WD surface.

`Dwarf nova' is an umbrella term, covering systems with similar, yet subtly different, outbursts as observed at optical wavelengths (U~Gem, SU~UMa, ER~UMa and Z~Cam; see chapter by Webb). However, the X-rays (in the non-magnetic systems) are always produced in the same way, namely through shocks within the boundary layer (BL) between the accretion disc and WD surface \cite{patterson85a, patterson85b}. The Keplerian velocity of the inner disc is much faster than the rotation of the WD (in the range of $\sim$~3000~km~s$^{-1}$, compared with $\sim$~300~km~s$^{-1}$ for the WD surface velocity), so the accreting material has to slow down over a short distance. This lost kinetic energy heats the BL to X-ray temperatures, leading to it emitting up to half the total luminosity of the system, with the remaining part released in the disc \citep[e.g.][]{lyndenbell74}. 

The type of X-ray emission from the BL is dependent on the accretion rate. During quiescence, the rate of mass transfer through the BL is relatively low. In this case, optically-thin emission occurs, as the in-falling material collides with the WD surface forming hard X-ray shocks \cite{patterson85a}. However, this is not the end of the story. Hot, relatively low-density gas such as this does not radiate efficiently: cooling should mainly occur through bremsstrahlung (free-free) radiation, which relies on interactions between charged particles, of which there are few in this rarefied material. In the case of particularly low density, the hot gas therefore expands, leading to even fewer opportunities for collisions/radiation, and hence expanding even further. This process leads to the BL puffing up into a diffuse, hot, X-ray-emitting corona \citep[e.g.][]{meyer94}. As material flows towards the WD, it releases gravitational potential energy, continuing to heat the corona, which itself conducts energy into the accretion disc. Evaporation from the disc serves to replenish gas lost from the corona -- and so the cycle repeats. Harder X-rays of up to 20~keV (see, e.g., \cite{warner} for relevant calculations)  will be formed in such a hot corona, whereas shocks with kT around a few keV will occur within the gas collapsing directly onto the WD surface, generated when the accretion rate, and hence density, is somewhat higher \cite{pringle79}.

During the DN outburst, however, the accretion rate, and therefore the amount of material in the BL, increases significantly. Thus, there are more particles to experience more collisions, and the BL can cool more readily. Under these conditions, no corona is formed, and the BL becomes optically-thick to X-rays. The shocked photons are absorbed and thermalised within the layer \cite{patterson85b}. Because of this, the harder X-ray emission during outburst is usually strongly suppressed in comparison to quiescence, with a corresponding increase in the extreme UV (EUV) and/or soft X-rays. This was first observed in detail for SS Cyg in 1996 by \cite{pete03}, combining data from \rxte\footnote{Rossi X-ray Timing Explorer} and \textit{EUVE}\footnote{Extreme Ultraviolet Explorer}, where the hard X-rays (\rxte\ was only sensitive to photons above 1.5~keV; no simultaneous soft X-ray data were collected) were seen to rise shortly after the start of the optical outburst, and then abruptly quench as the EUV emission took over. Towards the end of the optical outburst, the X-rays (and, slightly delayed again, the EUV) undergo another temporary rebrightening. The lag between the start of the rises at optical and X-ray energies is attributed to the propagation time of the heating wave moving through the disc to reach the BL, before being accreted, while the delay between the X-ray and UV increases implies the UV photons are the soft tail of the optically-thin, hard X-ray component. Despite these plausible explanations for the temporary X-ray flares at the start and end of the outburst, \cite{fertig11} point out that most DNe do not actually show these features, though \cite{byckling09, neustroev18} provide two other examples where such brief X-ray increases at either end of the outburst are indeed seen.

\begin{figure}
\begin{center}

  \includegraphics[clip, angle=-90, width=10cm]{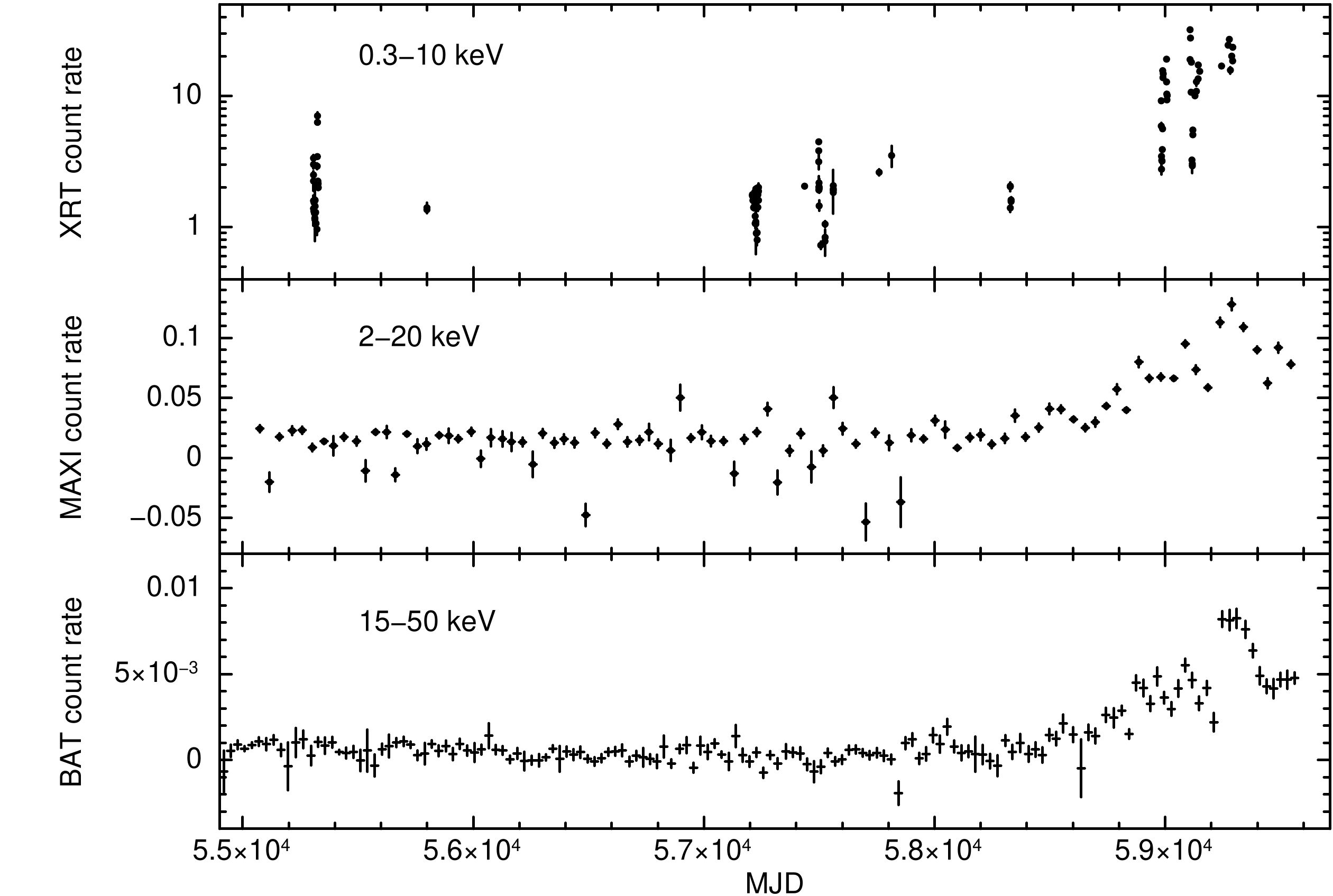}
  \includegraphics[clip, angle=-90, width=10cm]{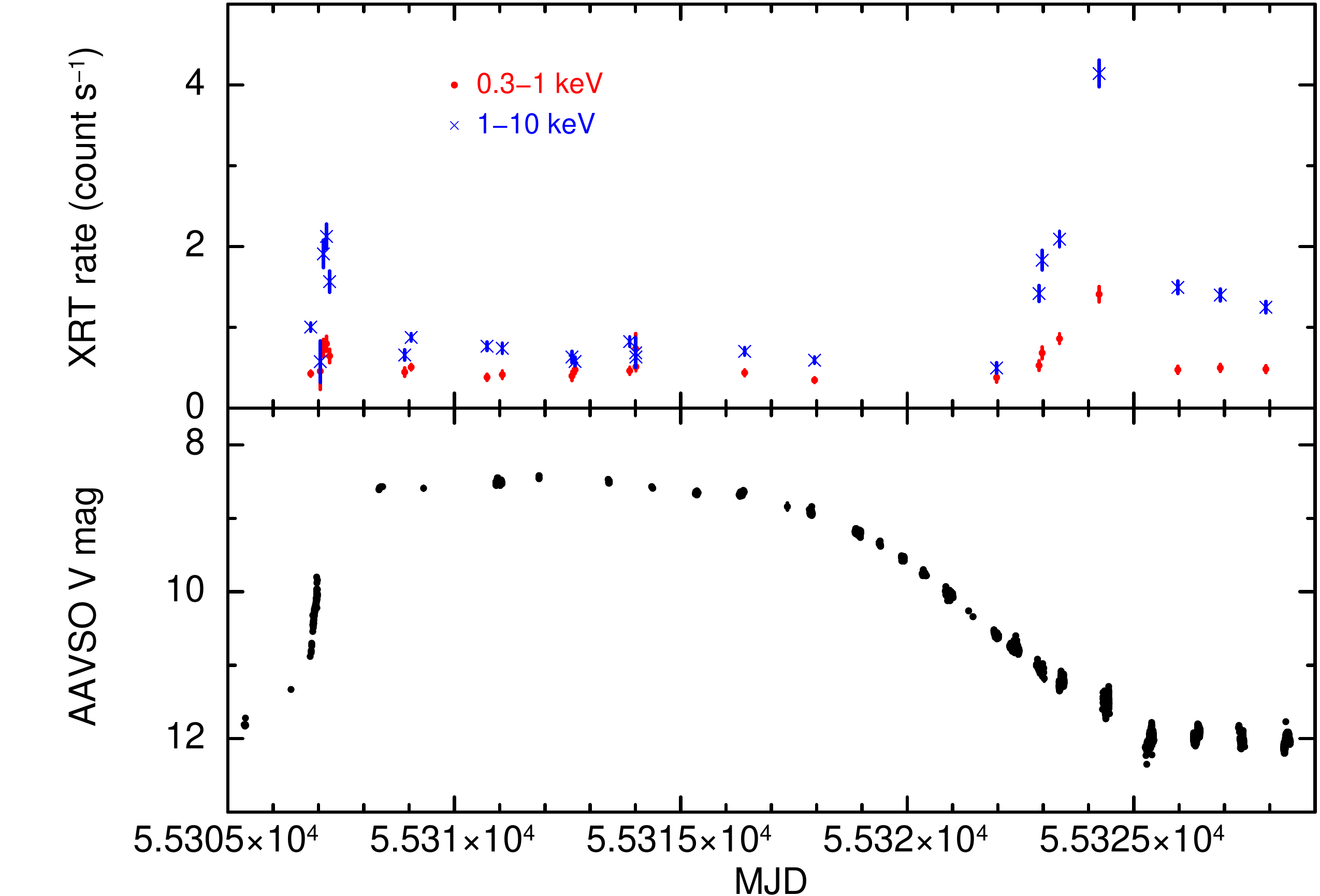}
\caption{Top: X-ray observations of SS Cyg from 2009 to 2021. The upper panel shows data from pointed observations with the narrow-field XRT onboard {\it Swift}, which is why there are large gaps in the coverage; the middle and lower panels show data from {\it MAXI} and {\it Swift}-BAT, which constantly survey areas of the sky. Bottom: The 2010-April outburst of SS Cyg observed over the X-ray and optical bands. The upper panel shows the soft (0.3-1 keV) and hard (1-10 keV) X-ray light curves overplotted. The time axis is Modified Julian Date.}
\label{sscyg}
\end{center}
\end{figure}

Fig.~\ref{sscyg} shows long-term X-ray observations of SS~Cyg in the top plot, covering 2009--2021; SS Cyg undergoes a DN outburst approximately every 50 d. Interestingly, since 2019 the hard X-rays ($>$~10~keV) have mainly shown a brightening trend. The bottom plot shows the outburst of SS Cyg in 2010 April, observed in X-rays\footnote{Light curves produced using the online {\it Swift} XRT product generator at \url{https://www.swift.ac.uk/user\_objects}} and optical\footnote{Data taken from AAVSO: American Association of Variable Star Observers.} \cite[see also][]{russell16}. The UV source was too bright for the {\it Swift} UVOT\footnote{UV/Optical Telescope} to observe usefully. The rise and rapid switch-off of the X-rays shortly following the start of the optical outburst, and again, towards the end, can be seen, with the majority of the increase above 1~keV. During the main body of the optical outburst, the X-ray spectrum is softer.

While the majority of DNe are X-ray fainter during the outburst, some instead show an increase in X-ray luminosity during this time, with an appearance, or strengthening, of a harder X-ray component. This is still not fully understood -- and neither, indeed, is the reason why there are residual harder, optically thin X-rays detected during outburst, when the BL is expected to be completely optically thick (see further discussion in \cite{Mukai-2017}). 

An example of a DN exhibiting unexplained phenomena is
SSS~J122221.7-311525, a WZ Sge-type DN (the subset of the SU~UMa DNe which show only superoutbursts), which was followed in detail throughout one such superoutburst by {\it Swift} \cite{neustroev18}. This source showed a significantly higher X-ray luminosity during outburst than at quiescence (displaying an optically-thin spectrum, with no sign of the expected soft component; as mentioned by \cite{byckling09}, however, even a small spectral change could lead to the soft emission falling below the XRT bandpass, emerging in the EUV band instead), and, interestingly, the X-ray flux varied abruptly at the same time as the optical superhump behaviour changed (an effect also seen in observations of GW Lib presented in the same paper). This suggests a link between the inner region (from where the X-rays are emitted) and further out in the disc (from where the superhumps arise), indicating that the accretion disc models for such systems are incomplete.

Note that, while the majority of DNe are non-magnetic systems, as considered in this section, some magnetic CVs (such as GK~Per; \cite{evans09}) also show this kind of outburst. However, as discussed in Section~\ref{sec:mag}, there is no boundary layer in magnetic CV systems, since the inner part of the accretion disc is disrupted. In such cases, most of the energy is released in the stand-off shock close to the polar regions, in the form of optically-thin plasma. This emission is likely only weakly dependent on the accretion rate \cite{hameury17}, and can be used to constrain the magnetic field in these objects \cite{Suleimanov-2016}.

\section{Combination Novae}

As an aside, we note that there are also `combination novae': sources which show both dwarf and classical nova characteristics. As for the other populations of accreting WDs discussed in this chapter, the classification of the sources is not based on the X-ray characteristics, but rather the optical wavelengths. Combination novae undergo an event which starts off as an accretion disc instability, leading to enough of the disc material being accreted onto the WD to trigger the thermonuclear shell burning and mass ejection typically expected in a classical nova. Such an outburst was first reported by \cite{sok06} in the symbiotic star Z And, where the authors established that, while the start of the event resembled a DN-like disc instability, the total eruption was too energetic to be purely accretion driven. X-ray observations by \chandra\ and \xmm\ showed soft spectra, with the majority of the photons below 2~keV, although there were relatively few source counts collected in two of the three observations. \cite{sok06} suggest that many classical symbiotic outbursts may actually fall under this heading of combination novae, and \cite{bollimpalli18} consider that the recurrent nova eruptions in RS Oph could be triggered by disc instabilities (though they believe that the Z And combination nova event is more likely to have been triggered by increased mass transfer from the giant companion, rather than a disc instability).

\section{Nova-like Variables}

Nova-like systems are non-magnetic CVs in which the mass-transfer rate is high enough ($\dot{M}\geq10^{-9}$ $\msun$ yr$^{-1}$) to sustain the disc in a perpetually hot state. Because the disc is stable, they will not show dwarf nova outbursts (that is, they are termed non-eruptive systems), instead maintaining an outburst-like steady state. 

At such a high $\dot{M}$, one would expect the emission mechanisms to be similar to DNe in outburst, i.e. a BL with an optically thick, soft X-ray emitting component. In the VY\,Scl systems -- nova-likes that transition between high and low states due to varying mass accretion rate; they are also known as anti-dwarf novae -- at lower $\dot{M}$ the BL is expected to be optically thin and emit hard X-rays. However, there are few high-quality X-ray observations of nova-likes that allow us truly to determine the nature of the X-rays. Deep \xmm\ observations of the high-inclination nova-like UX\,UMa found an X-ray spectrum with two components. The soft X-ray emission was uneclipsed, suggesting an extended origin. The hard component, conversely, was eclipsed, such that it must originate close to the WD surface within the BL \citep{Pratt-2004}, questioning the notion that high $\dot{M}$ BLs are strong soft X-ray emitters. Similarly, \citep{Balman-2014} found that, in three high-state nova-like systems, the X-ray spectra were dominated by optically-thin emission with maximum temperatures in the range 21--50 keV, and postulated that the BL may be merged with an advection dominated accretion flow (ADAF) and/or an X-ray emitting `corona' as in X-ray binaries\footnote{Note that, while accreting binary systems containing WDs are referred to as CVs, those with neutron stars or black holes are termed X-ray binaries}. The few deep X-ray observations of nova-likes seem to point to multiple origins for the X-ray emission, and appear to have posed more questions about the emission mechanisms than they have answered.

\section{Persistent super-soft sources}

Besides novae which may pass through a transient super-soft phase, there is also a population of persistent (or, at least, very long-lived) luminous super-soft sources. A number of strong X-ray sources in the Large Magellanic Cloud (LMC) were analysed by \cite{vandenheuvel92}, who found that the emission could be explained by \hbox{(quasi-)} steady hydrogen burning on WDs when the accretion rate is sufficiently high ($\sim$~10$^{-7}$~$\msun$~$\rm yr^{-1}$) -- the so-called `close binary super-soft source' (CBSS) model \cite[see also][]{kahabka97}. The high accretion rate in such systems is thought to arise due to the mass-losing companion star being more massive than the WD (whereas, in the case of a nova, the WD has the larger mass). Alternatively, in some steady super-soft systems the WD may accrete via wind-driven mass transfer from a low-mass companion which is being irradiated by the WD itself \citep{vanteeseling98}.

\begin{figure}[t]
\begin{center}
  \includegraphics[clip, width=8cm, angle=-90]{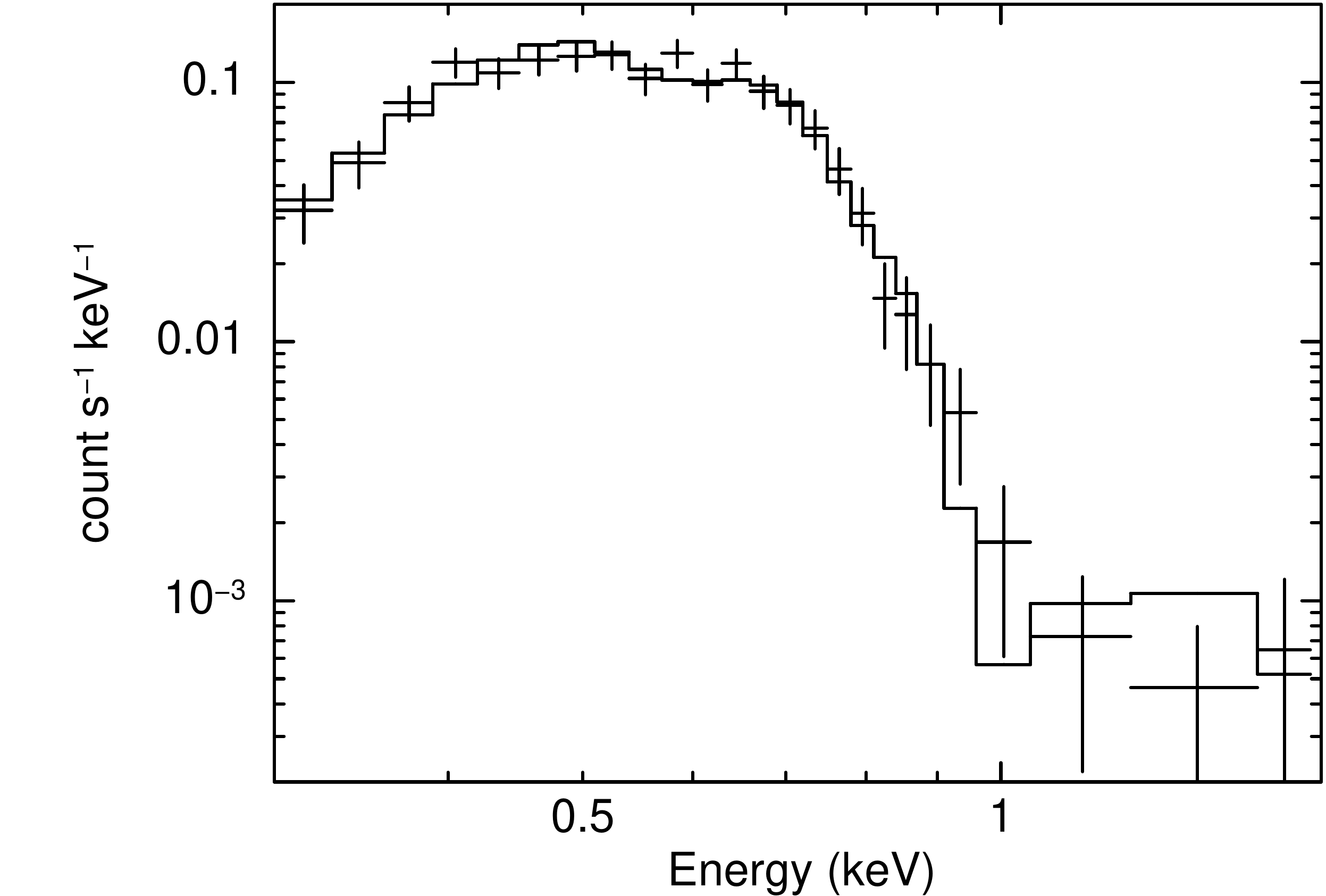}
\caption{{\it Swift}-XRT spectrum of persistent super-soft source CAL 87. Almost all the detected counts are soft, and, as for novae in their SSS phase, the spectrum can be parameterised with either a BB plus edges (as shown here), or a stellar atmosphere model.}
\label{perSSSspec}
\end{center}
\end{figure}

Despite the `persistent' in the name, these sources do pass through X-ray faint states. It has been speculated \citep{reinsch00} that these `off' intervals are caused by the accretion rate increasing, causing the WD photosphere to expand, thus shifting the peak of the emission out of the X-ray band and into the EUV. When the photosphere shrinks back down, the source will rebrighten in the X-rays. This is similar to the changes measured in the SSS temperatures in novae described in Section~\ref{nova:lc}, which may help to explain the high-amplitude flux variability sometimes seen.

\begin{figure}[t]
\begin{center}
  \includegraphics[clip, width=10cm]{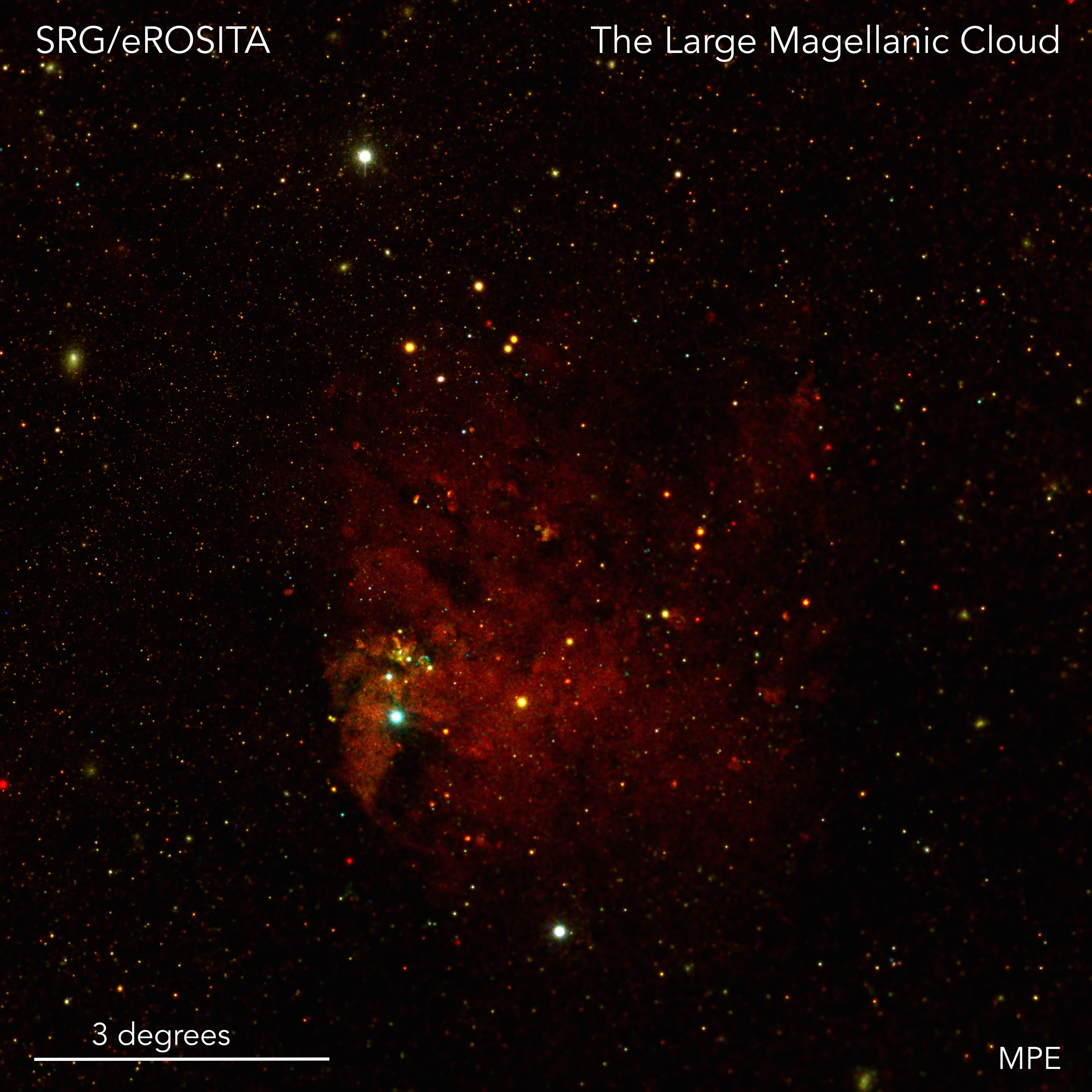}
\caption{{\it eROSITA} false colour image of the Large Magellanic Cloud. Soft sources appear red, and harder ones blue. Credit: Frank Haberl, Chandreyee Maitra (MPE). }
\label{lmc}
\end{center}
\end{figure}

Fig.~\ref{perSSSspec} plots the {\it Swift}-XRT spectrum of CAL 87, one of the persistent SSS in the LMC, highlighting the softness of the emission, with very few counts above 1~keV; it looks very similar to the RS~Oph SSS spectra shown in Fig.~\ref{var}. Fig.~\ref{lmc} shows an {\it eROSITA} image of the LMC. In RGB false colour images such as this, softer sources (e.g., persistent SSS or novae passing through a super-soft phase) appear redder, while harder sources are bluer.

\section{BeWD systems}

Be X-ray binaries are a type of High Mass X-ray Binary (HMXB) comprising a Be-type star (that is, a B-star which shows emission lines) and (usually) a neutron star; a large number of these systems are known. BeWD systems, where the compact object is, instead, a WD, have been predicted to be even more numerous than their higher mass BeNS counterparts \citep{raguzova01}, though, to date, only a handful have been detected (see summary in \cite{coe20}).  The \swift\ Small Magellanic Cloud Survey (known as S-CUBED; \cite{kennea18}) is starting to discover more of these objects \citep{coe20, kennea21} when they undergo an outburst and flare up in X-rays, due to enhanced accretion. The X-ray emission of a BeWD system is soft, showing a spectrum similar to the SSS discussed above, although the X-rays are produced through shocked radial accretion, not nuclear burning \cite{kennea21}.

\section{Symbiotic stars}
Symbiotic binaries are those where a WD\footnote{There are also some neutron star symbiotic systems, but we do not consider those here.} accretes from an RG companion. In some cases, where the WD mass is high enough and sufficient mass transferred, these systems can erupt as recurrent novae -- RS Oph, for example, as mentioned above.

As discussed by \cite{luna13}, when modifying and expanding an original classification scheme based on {\it ROSAT}\footnote{R{\" o}ntgensatellit} data from \citep{muerset97}, WD symbiotics can be split into groups which show only super-soft emission (i.e. $<$~1~keV, but mainly $<$~0.4~keV); photons over the full {\it ROSAT} bandpass (up to 2.4~keV); highly absorbed, harder X-rays (detectable above 2.4~keV -- up to tens of keV in some cases; e.g. \cite{kennea09}); and sources which show both soft and hard components. 
The production mechanisms for these different energies of X-ray photons are similar to those for other accreting WD systems covered earlier in this chapter. The super-soft emission is likely caused by hydrogen burning on the WD surface \cite{orio07} (which may be a residual from previous nova outbursts, e.g. \cite{ramsay16}), while the not-quite-so-soft emission is thought to be generated through the low-velocity shock collision between winds from the shell-burning WD and the RG \cite{muerset97}. As for DNe, the harder, optically-thin X-rays are most probably accretion powered, coming from the BL between the WD and accretion disc \cite{kennea09}.

In the case of a small number of symbiotic stars, extended, spatially-resolved X-ray emission has also been identified, and interpreted as being related to the jets detected at radio wavelengths  \citep[e.g.][] {nichols07}.

\section{Oddballs}
\label{odd}

While most super-soft sources are typically thought to be caused by nuclear burning, observations of a source called ASASSN-16oh showed a super-soft X-ray source with no accompanying sign in the optical light curve that nuclear fusion had turned on, together with lines in the contemporaneous optical spectra clearly in agreement with a disc origin (that is, similar to a DN outburst) rather than nova ejecta; the X-rays also had to be coming from only a small fraction of the WD surface \citep{tom19}. The authors suggest that the emission came from the sudden accretion of a spreading layer in the form of a belt around the WD \citep[see][]{kip78}, rather than a classical BL.

An alternative explanation for ASASSN-16oh was put forth by \cite{hillman19}, whereby the source underwent a TNR, but without the usual accompanying mass ejection; models have shown that such events should be feasible \citep[][and references therein]{hillman16}. In this model, the low luminosity compared with the output from `normal' novae is caused by an optically-thick accretion disc obscuring the WD surface from our view.

ASASSN-16oh is therefore an example of an accreting WD system where the provenance of the X-rays is not obvious.

\section{Magnetic Cataclysmic Variables }
\label{sec:mag}

\begin{figure}
    \centering
    \includegraphics[width=\textwidth, angle=0]{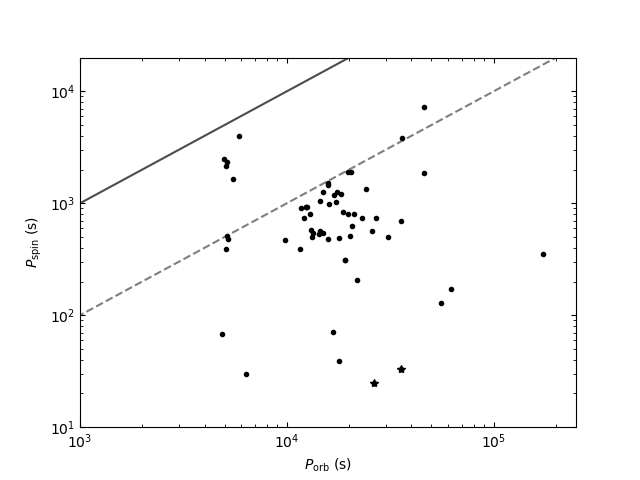}
    \caption{Diagram showing the firm \pspin\ and \porb\ measurements of IPs taken from Koji Mukai's catalogue at  \url{https://asd.gsfc.nasa.gov/Koji.Mukai/iphome/iphome.html}. The solid diagonal line shows the line of equality, where \pspin = \porb, i.e. where the polars lie. The dashed diagonal line represents the \pspin=0.1\porb\ equality determined to be the upper limit for partial disc formation in IPs \citep{King-1991}. IPs above this line are either severely out of spin equilibrium, or the accretion flow is unlike the Keplerian disc we typically assume is present in IPs (see Section \ref{sec:IP_LCs}). The stars highlight the two probable `propeller' systems, AE\,Aqr and LAMOST\,J024048.51$+$195226.9 (see Section \ref{sec:AEAqr}).}
    \label{fig:spin-orb}
\end{figure}

Magnetic cataclysmic variables (mCVs) are so named for the strong magnetic field ($B$ $\sim$ 10$^6$ -- 10$^8$ G) of the WD primary \citep{Cropper-1990,Patterson-1994}. The magnetic field disrupts the accretion disc, forcing material to flow along the field lines on to the magnetic poles of the WD. We can generally split mCVs into two main sub-classes: polars and intermediate polars (IPs). In polars, the magnetic field strength, $B$, of the WD is strong enough such that matter flows along the field lines from the donor without forming a disc. However, in IPs, only the innermost regions of the disc are disrupted, and they typically exhibit a residual accretion disc which terminates at the magnetospheric boundary, \rmag. Aside from their magnetic field strengths, the two sub-classes are easily observationally distinguished through the ratio of the white dwarf's rotational period (\pspin) to the binary orbital period (\porb). In polars, the magnetic field is strong enough to synchronize the two such that \pspin=\porb, but in IPs typically \pspin~$<<$~\porb\footnote{There is a small population of so-called `asynchronous' polars (APs) which are very similar to polars, but their \pspin\ and \porb\ differ by a factor of $\sim1\%$ \citep{Campbell-1999}. In some cases, this asynchronicity may have been caused by a nova explosion in the system.} (Fig.~\ref{fig:spin-orb}).

\begin{figure}
    \centering
    \includegraphics[width=\textwidth]{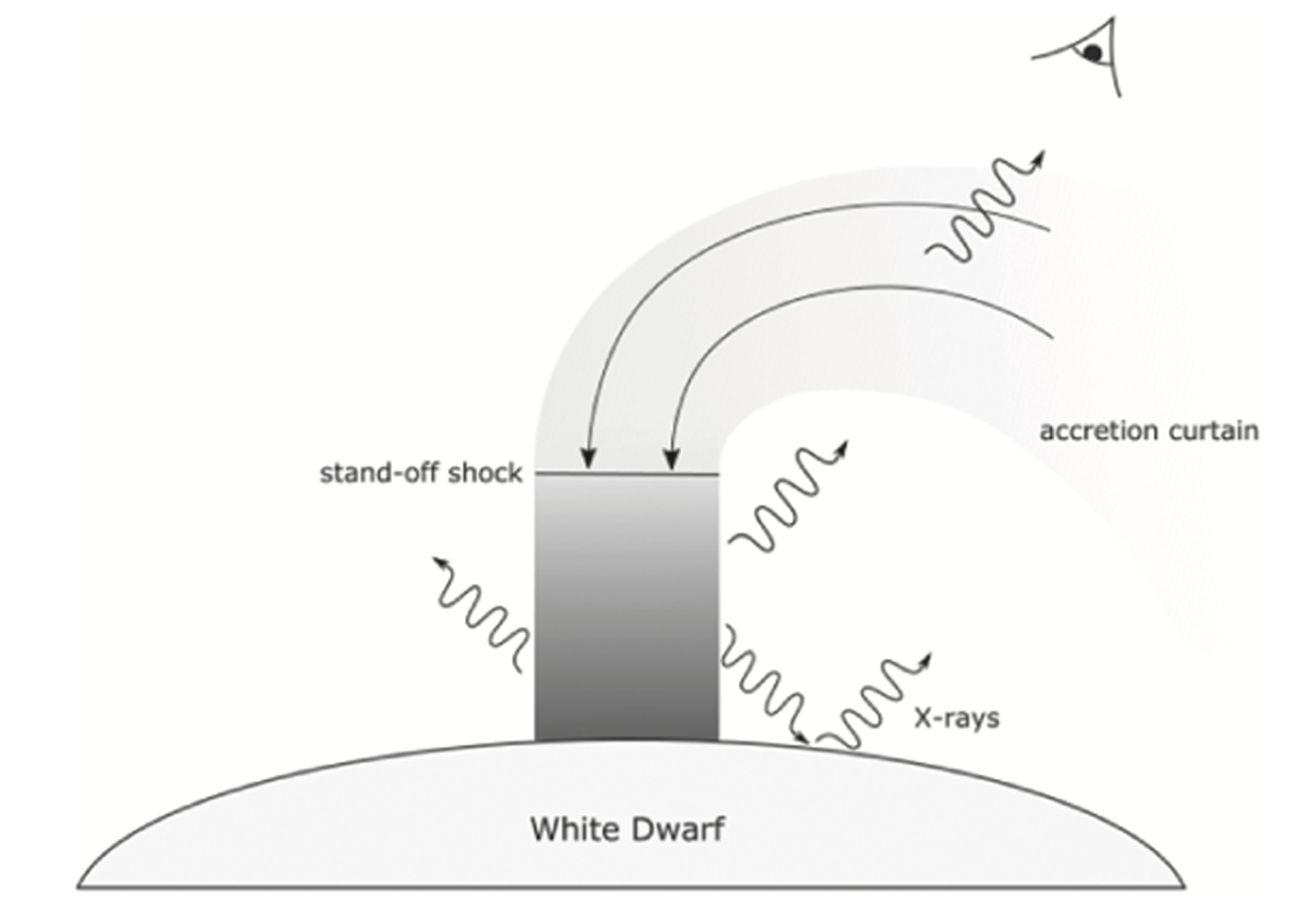}
    \caption{A schematic of the geometry of an mCV and the associated X-ray emission regions. The primary source of hard X-ray photons is the cooling standing shock above the magnetic pole of the WD, and these photons can be absorbed by the accretion curtains or reflected off the WD surface (see e.g. Section \ref{sec:refl}). Reproduced with permission from \citep{Hailey-2016}.}
    \label{fig:Hailey_IP}
\end{figure}

Regardless of magnetic field strength, all mCVs are strong X-ray emitters and in this section we detail the emission processes responsible for the observed behaviour at high energies. Similar to DNe and other classes of non-magnetic accreting WDs, a large fraction of the X-ray emission we observe originates in the shock that is formed as accreted material approaches the surface of the WD. However, instead of forming a BL at the interface between the Keplerian disc and the surface of the WD proper (as in non-magnetic CVs), the shock manifests as a so-called `accretion column' at the magnetic poles of the WD (see Fig. \ref{fig:Hailey_IP} for a schematic). Standing shocks in mCVs are heated to temperatures \kts~$\gtrsim~10$ keV and cool through thermal bremsstrahlung as the post-shock material settles onto the surface of the WD, emitting hard X-rays in the process. The physics of the post-shock region (PSR) were first explored by \citep{Aizu-1973} in the context of the (non-magnetic) CV SS\,Cyg, but theories were subsequently developed over several decades \citep[e.g.][]{Fabian-1976,Wu-1994a,Wu-1994b,Cropper-1999,Suleimanov-2016} to form the basis of our understanding today.

\subsection{X-ray spectra of mCVs}
\label{sec:mCVspectra}

The typical physical picture of (hard) X-ray emission in mCVs is that matter will fall on to the WD from a distance that can be approximated as $\infty$, such that the in-falling material reaches the free-fall velocity $v_{\rm ff}$:

\begin{equation}
    v_{\rm ff}=\sqrt{\frac{2GM_{\rm WD}}{R_{\rm WD}}}
    \label{eq:freefall}
\end{equation}

\noindent where \mwd\ and \rwd\ are the mass and radius of the WD, respectively. In IPs, it should be noted that one must take into account the inner radius of the accretion disc at \rmag, such that the in-falling material does not necessarily fall from $\infty$, and thus Equation \ref{eq:freefall} must be multiplied by a $\sqrt{1-r_{\rm m}^{-1}}$ term, where $r_{\rm m}$ is the magnetospheric radius relative to the WD radius, $r_{\rm m}=\frac{R_{\rm m}}{R_{\rm WD}}$ \citep{Suleimanov-2016}. The accreted matter forms an adiabatic shock above the surface of the WD, crossing the shock at a velocity $\frac{1}{4}v_{\rm ff}$ as implied by the Rankine-Hugoniot relations in the strong shock limit. Thus, the resultant conversion of kinetic to thermal energy takes the form:

\begin{equation}
    kT_{\rm shock} = \frac{3}{16}\mu m_{\rm H} v_{\rm ff}^2
    \label{eq:temp}
\end{equation}

\noindent where $\mu$ is the mean molecular weight of the accreted gas, which is typically assumed to be $\approx0.62$ (a fully ionized plasma with solar composition), and $m_{\rm H}$ is the mass of a proton. Thus, the maximum shock temperature $kT_{\rm shock}$ (i.e. at the top of the shock) is directly related to the mass and radius of the WD:

\begin{equation}
    kT_{\rm shock} = \frac{3}{8}\frac{\mu m_{\rm H} G M_{\rm WD}}{R_{\rm WD}}.
    \label{eq:temp_MR}
\end{equation}

\noindent Again, other factors such as matter falling from \rmag, as well as the height of the PSR itself can be taken into account here \citep[see][]{Suleimanov-2016,Suleimanov-2019} but, generally, Equation \ref{eq:temp_MR} illustrates the dominant thermal process in the shock. 

The mass--radius relationship of WDs has been well-studied for some time now \citep[see, e.g.][]{Nauenberg-1972}, such that we can show that, for a reasonable range of masses, \mwd~=~0.6--1.2 $\msun$, we might expect $kT_{\rm shock}$~=~21--90~keV. As the material settles onto the surface of the WD, it must cool to the WD photosphere temperature via optically thin bremsstrahlung emission. As a result, the hard X-ray spectrum of mCVs, though in reality a sum of multi-temperature plasma emission, can often be well approximated by a thermal bremsstrahlung model.

\cite{Mukai-2003} showed that IP spectra can also be well described by the isobaric radiative `cooling flow' model. Originally developed to describe the X-ray emission in clusters of galaxies \citep{Mushotzky-1988}, the cooling flow model has been used successfully to fit the X-ray spectra of IPs across a broad energy band \citep[e.g.][]{Mukai-2015,Luna-2018}.

As one might expect from a hot, multi-temperature plasma, at high spectral resolutions (such as that offered by the HETG onboard the \chandra\ X-ray telescope), emission lines from highly ionized metals become prominent in the X-ray spectra of mCVs. While H- and He-like ions of, e.g., Si, Mg and Ne are often present \citep{Ishida-1994,Mukai-2003}, as well as the entire L-shell complex of Fe, the most prominent line emission comes from Fe in the 6-7 keV energy range. The 6-7 keV Fe complex can be split into neutral Fe K$\alpha$ at 6.4 keV, He-like Fe XXV at 6.7 keV and H-like Fe XXVI at 6.97 keV. Neutral Fe K$\alpha$ is expected through reprocessing of X-rays by the absorber surrounding the PSR. However, \cite{Ezuka-1999} showed that a non-negligible fraction of Fe K$\alpha$ originates from reflection of PSR X-ray photons from the WD surface (see Section \ref{sec:refl}).

Given the relationship between the WD mass, radius and shock temperature (Equation~\ref{eq:temp_MR}), the X-ray spectrum of mCVs can actually be used to derive WD masses, as will be discussed in more detail in Section~\ref{sec:mass}.

\subsubsection{Cyclotron cooling in polars}
\label{sec:cyclotron}

While the bremsstrahlung interpretation described above works well for IPs, in the presence of high magnetic fields like those in polars, one must consider the effects of cyclotron cooling on the X-ray spectrum. The importance of cyclotron cooling depends heavily on the specific mass accretion rate $\dot{m}$ and the magnetic field strength $B$. In IPs, the specific accretion rate is high, $\dot{m}>1$ g s$^{-1}$ cm$^{-2}$, and the magnetic field strength is relatively low, $B\sim10^6$ G, such that cyclotron cooling can effectively be ignored. However in polars, $B$ is typically much higher ($>10^7$ G), and their specific accretion rates are low. In this regime, cyclotron cooling suppresses the bremsstrahlung cooling \citep{Woelk-1996}, though bremsstrahlung losses can never be completely ignored \citep{Wu-1994b}. Cyclotron emission causes the shock height to reduce by 1-2 orders of magnitude and reduces the electron temperature of the PSR by an order of magnitude \citep{Wu-1994b,Woelk-1996} compared to the IP case.

\subsubsection{Reflection}
\label{sec:refl}

\begin{figure}
    \centering
    \includegraphics[width=\textwidth]{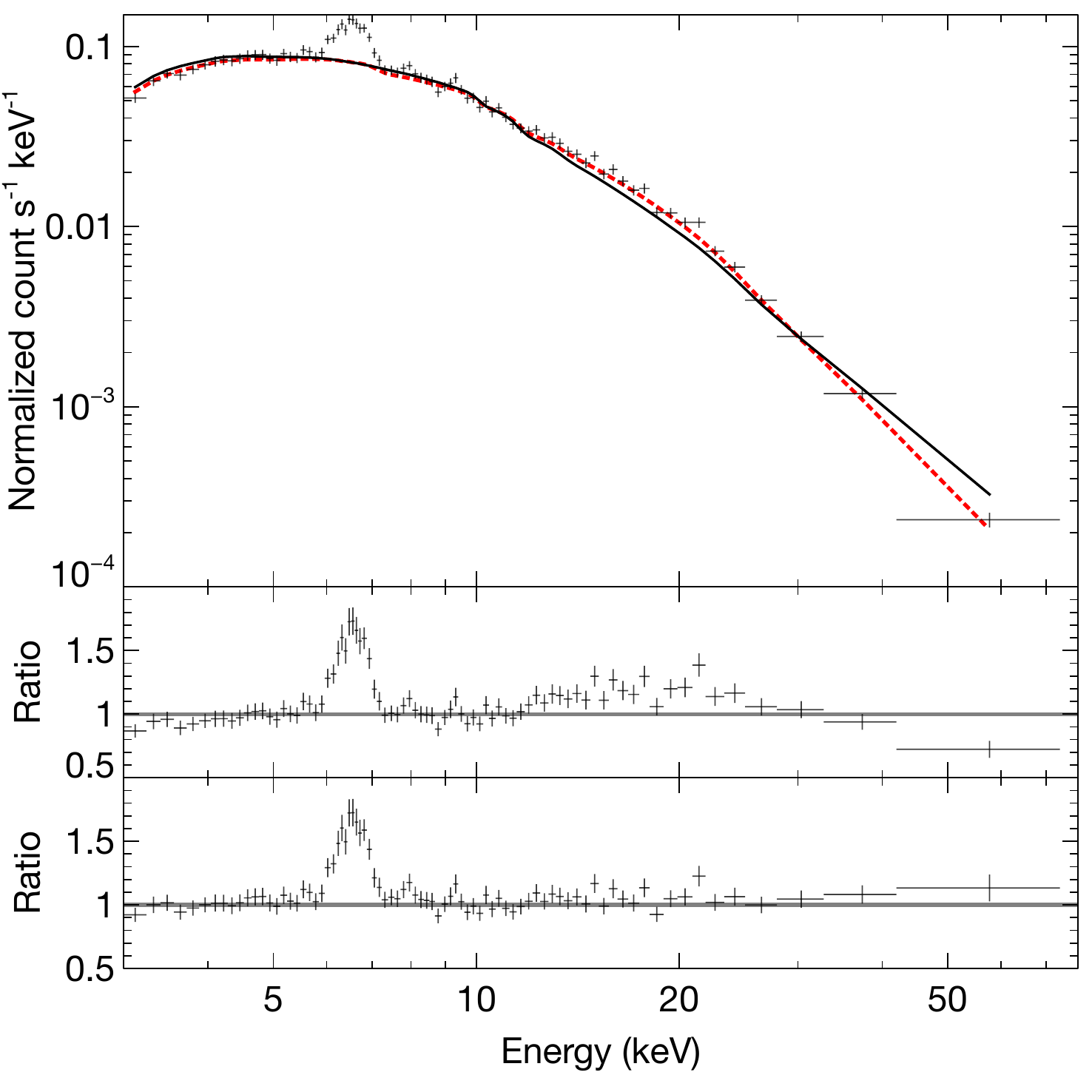}
    \caption{The 3--78 keV \nustar\ spectrum of the IP NY\,Lup. The top panel shows the X-ray spectrum and a best fit bremsstrahlung model (black solid line) and a bremsstrahlung model modified with a reflection component (red dashed line). The middle panel shows the data/model ratio of the bremsstrahlung-only fit; note the curvature of the spectrum at the high-energy end. The bottom panel shows the data/model of the reflection fit, showing that the reflection model accounts for the curvature (or `Compton hump'). For clarity we show data from only one of the \nustar\ focal plane modules. See \cite{Mukai-2015} for more details on the data behind the figure.}
    \label{fig:NYLup_reflection}
\end{figure}

With X-rays being produced so close to the surface of the WD in mCVs, one might expect $\sim$half of the emitted radiation to be directed toward the WD and reflected back to the observer. X-ray reflection in accreting compact objects manifests in the X-ray spectrum  as a broad Compton `hump' in the 10--30 keV range, and a neutral Fe K$\alpha$ line at 6.4 keV \citep{George-1991}, and is often detected in X-ray binaries and AGN as a result of photons being reflected from the accretion disc. In the case of mCVs, the importance of reflection has been implied due to the presence of the Fe K$\alpha$ line \citep{Ezuka-1999}. However, with the exception of the prototypical polar AM Her \citep{Rothschild-1981} and EF Eri \citep{Done-1995}, detections of the Compton hump were originally difficult to come by. However, \cite{Mukai-2015} utilized the hard X-ray capabilities of the then-newly-launched \nustar\ observatory unambiguously to detect the reflection continuum in IPs for the first time (see Fig. \ref{fig:NYLup_reflection}).

\cite{Mukai-2015} were also able to measure the amplitude of the reflection component, which provides a way of estimating the height of the shock. For reference, a reflection amplitude of unity implies that the X-ray emitter (i.e. the shock) is just above the surface of the reflector (the WD). A deviation from 1 would therefore imply a non-negligible shock height. This was found to be the case for V709\,Cas, where the measured reflection amplitude $<0.6$ can be explained by a shock $\sim0.2$\rwd\ above the surface of the WD \citep{Mukai-2015}. The shock height in IPs can also be probed by studying the spin modulation in the X-ray light curves (see Section \ref{sec:IP_LCs}). It is important to note here a caveat to measuring reflection in mCVs, in that the modification of the spectrum due to reflection can (at least partly) be mimicked by partial absorption, for instance by the material in the accretion curtain. Such partial covering is, in fact, a common phenomenon in IP spectra (see also Section \ref{sec:IP_LCs}). Further studies with high resolution broadband instruments like \nustar\ are, therefore, necessary to understand fully the role of reflection in the IP population as a whole.

\subsubsection{The soft component of mCVs}
\label{sec:soft_component}

Since the early days of X-ray astronomy, it has been noted that polars are strong sources of soft X-ray emission \citep[e.g.][]{Lamb-1979}. If one assumes that $\sim50$\% of the X-rays from the shock are directed towards the WD (see Section \ref{sec:refl}), then soft X-ray (as well as EUV) emission from absorbed and reprocessed photons is not a surprise \citep{Lamb-1979}. However, it was quickly established that polars in fact exhibited an excess of soft X-ray emission compared to that expected from simple reprocessing of bremsstrahlung and cyclotron photons \citep[see e.g.][]{Ramsay-1994,Beuermann-1995}, with comparatively few examples of soft X-ray emission from IPs where similar physics might be expected. \cite{Kuijpers-1982} suggested that a soft X-ray excess could in fact occur through dense blobs of material penetrating the WD photosphere without forming a shock, instead heating the photosphere and producing a distinct low-energy BB component. Additionally, in the low-$\dot{m}$, high-$B$ regime of polars, the photosphere may be heated directly by a bombardment of fast ions \citep[e.g.][]{Fischer-2001}.

The soft X-ray excess was so prominent in early X-ray studies of polars that it was thought of as a defining characteristic of the subclass. However, \cite{Ramsay-2004a} investigated the X-ray spectra of 21 polars with the then-recently-launched \xmm\ satellite, in particular comparing with earlier work done with \rosat\ \citep{Ramsay-1994}. They found that the soft X-ray excess was not as prevalent as previously anticipated, finding a strong excess in only three systems, while at least six systems showed no evidence for a detectable soft component whatsoever. Meanwhile, \xmm\ was also instrumental in finding that a not-insignificant fraction of IPs also exhibited a soft BB component \citep{Evans-2007b}. Further studies have shown that $\sim30$\% of the entire IP population in fact have a soft component \citep{Bernardini-2017b}, so much so that polars can no longer be considered to be the `soft-emitting' class of mCVs.

The BB temperature of the soft component in IPs tends to be higher than is typically seen in polars, with inferred BB emission regions much smaller than those in polars. Such high temperatures (usually $>$50 eV)  would be locally super-Eddington if the soft component originates from the surface of the WD. Further adding to the puzzle is the fact that spectral fits do not support an alternative hypothesis that the soft component occurs in the coolest part of the PSR \citep{Bernardini-2017b}. \cite{deMartino-2020} theorize that perhaps there is a temperature gradient over the polar cap and we are only measuring the innermost, hottest regions. In addition, the shape of the accretion region in IPs may play a role in how we interpret the soft component.

\subsection{Masses of white dwarfs in mCVs}
\label{sec:mass}

CVs are binary systems, which paves the way for dynamical mass determination, via Kepler's Third Law, of WDs through optical spectroscopy. However, for the subclass of mCVs in particular, there is an alternative way to measure the mass of the WD through modelling of their X-ray spectra. This method requires no knowledge of the binary parameters, and bypasses the need for a measurement of the binary inclination, something which is extremely rare in non-eclipsing binaries. We showed in Equation \ref{eq:temp_MR} that the temperature of the shock can be directly related to the mass and radius of the WD. We show that, for a reasonable value of \mwd, we might expect \kts $\gtrsim$ 20 keV. This implies that, if we are in fact able to measure directly \kts\ itself, we could turn Equation \ref{eq:temp_MR} around and measure \mwd.

The power of the X-ray spectrum as a mass diagnostic was discussed as early as the 1980s in the context of AM Her, the prototypical polar \citep{Rothschild-1981}. \cite{Cropper-1998} presented an early mass survey of mCVs, modelling the PSR continuum spectra of 5 polars and 8 IPs using data from the {\em Ginga} satellite. They reported an average \avmwd = 0.96~$\msun$ for polars and \avmwd = 0.93~$\msun$ for IPs, while noting their model was limited by not including viewing-angle dependence on the absorption, or any variation in $\dot{m}$. Spectral modelling of mCVs prior to 2000 was also limited to energies $\lesssim20$ keV, as were probed by X-ray observatories of the time, despite the expected values of \kts\ implying a spectral turnover at much higher energies. As a result, some early studies focused instead on deriving \mwd\ through the measurement of the intensity ratios of H-like and He-like Fe lines, which are also a diagnostic of \kts\ \citep{Ezuka-1999}.

With the inclusion of hard X-ray instruments onboard X-ray telescopes such as \rxte\ and \suzaku, measurements of \mwd\ via the hard X-ray continuum became much more reliable. \cite{Suleimanov-2005} developed their own PSR model, applying it to the {\em RXTE} 3--100 keV spectra of 14 IPs and finding generally good agreement with the few IPs with independent optical measurements of \mwd. \cite{Yuasa-2010} applied the same model to 3--50 keV \suzaku\ spectra of 17 IPs, while also including the (resolved) 6--7 keV Fe emission complex in the model -- something of a hybrid of the continuum and emission line methods.\footnote{We note here that recent mass studies of mCVs have tended to focus mostly on IPs, not polars, as their spectra are relatively uncomplicated by cyclotron cooling (see Section \ref{sec:cyclotron}).}

However, though the rapid advancement of X-ray detectors heralded a significant improvement in mCV mass determination through the measurement of the hard X-ray continuum, the above studies still suffered from uncertain background. The hard X-ray instruments carried by \suzaku\ and \rxte\ (and all other X-ray missions prior to 2012) were non-imaging telescopes, meaning that the background had to be modelled rather than measured. Despite being strong hard X-ray emitters, IPs are still relatively faint in the hard X-ray band ($<10^{-10}$ \fluxcgs\ in the 14--195 keV band; \cite{Mukai-2017}). Thus, considering that a precise determination of \mwd\ depends heavily on accurately constraining the shape of a spectrum with \kts~$\gtrsim~20$ keV, good knowledge of the background is essential.

\begin{figure}[t!]
    \centering
    \includegraphics[width=\textwidth]{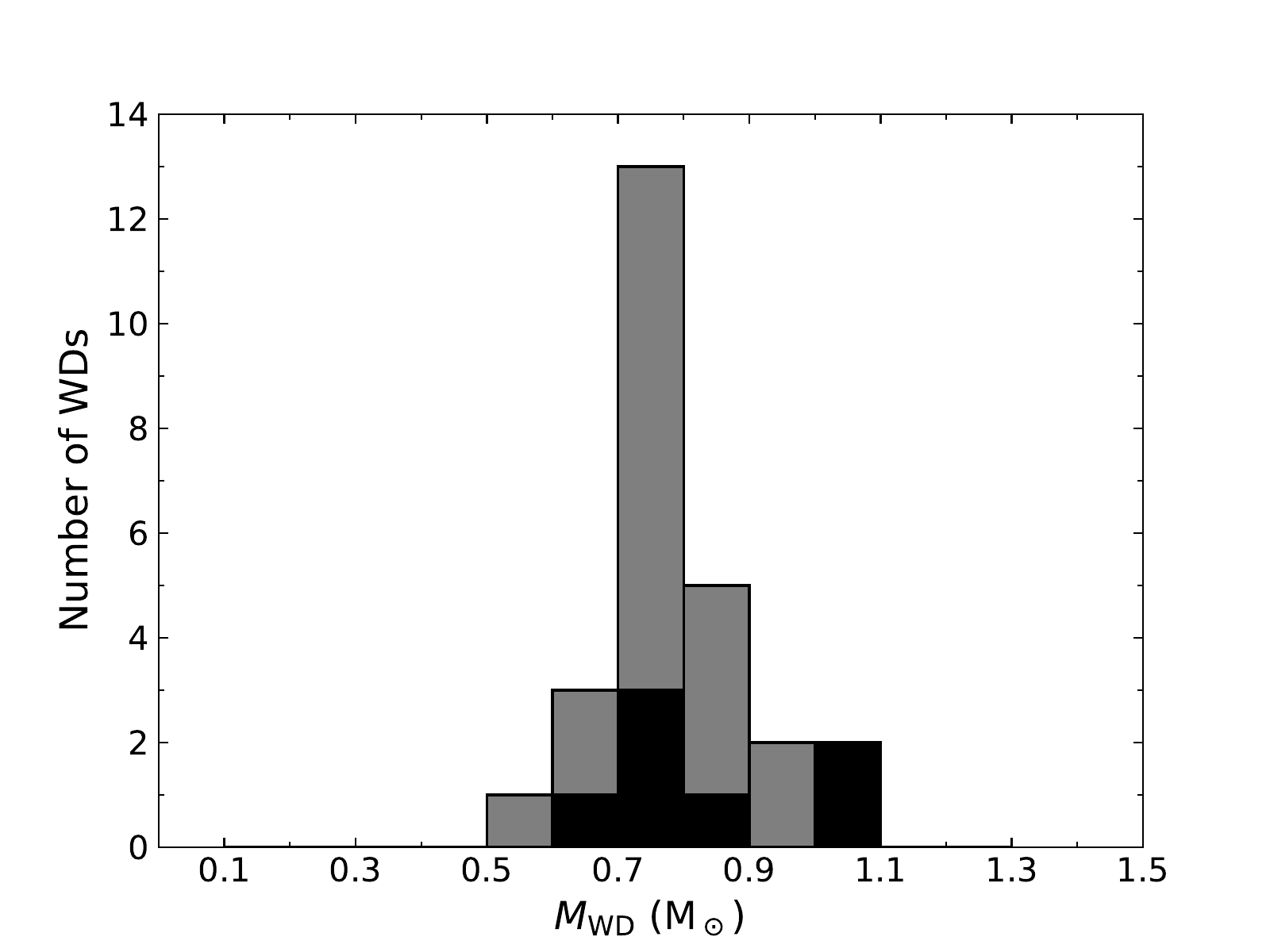}
    \caption{The mass distribution of WDs in mCVs, as measured by \cite{Shaw-2020b}, where \avmwd = 0.77 $\msun$. Highlighted in black are values of \mwd\ that had been derived from individual \nustar\ observations of seven IPs \citep{Suleimanov-2019} prior to the commencement of the dedicated Legacy survey that added a further 19 mCVs.}
    \label{fig:Legacy_masses}
\end{figure}

\nustar\ was launched in 2012, and is the first telescope capable of focusing X-rays in the 3--78 keV range. Being able to image at such energies means that background can be actually measured, rather than modelled, for each observation. \cite{Suleimanov-2016,Suleimanov-2019} further developed their continuum model of the PSR and applied it to 10 IPs observed by \nustar\footnote{The authors were also able to measure aperiodic variability in some IPs, providing an estimate of \rmag, which can affect the derived \mwd\ (see Section \ref{sec:IP_LCs}).} \citep[see also][]{Mukai-2015,Hailey-2016,Shaw-2018a}, deriving masses with a typical precision of $\sim3\%$. \cite{Shaw-2020b} brought the total of mCVs observed by \nustar\ to 26, measuring \avmwd = 0.77 $\msun$ (Fig.~\ref{fig:Legacy_masses}). \cite{Hailey-2016} also used \nustar\ data to conclude that the unresolved 20--40 keV X-ray emission emanating from the innermost 10 pc of the Milky Way is consistent with a population of IPs with \avmwd = 0.9 $\msun$.

\subsection{X-ray light curves of mCVs}
Though the spectra of mCVs have been discussed in detail in the preceding sections, in order to gain the most complete picture possible of the physics in these systems we must also consider their timing characteristics. mCVs are inherently variable objects, and in this section we will discuss the X-ray modulations observed in both polars and IPs.

\subsubsection{X-ray light curves of Polars}

When discussing the spin-modulated (recalling that \pspin=\porb) X-ray light curves of polars, \cite{Mason-1985} introduced the concept of `one-pole' and `two-pole' systems, referring to either (a) one accreting pole always being visible from Earth or (b) two poles being observed alternately as the WD spins, though only one is typically accreting. Naturally, this naming convention can lead to confusion, as one- or two-pole systems could also refer to the number of poles that are actually accreting, as well as the number that are visible from Earth. \cite{Mukai-1999} established the `upper pole' (pole on the same side of the orbital plane as Earth) and `lower pole' (opposite pole) nomenclature in part to remove this confusion (see Fig~\ref{fig:ip_curt} in Section~\ref{sec:IP_LCs} for example geometry). Nevertheless, the early studies of polars found that systems could be placed into two subcategories based on their phased X-ray light curves.

In the \cite{Mason-1985} \textit{EXOSAT}\footnote{European X-ray Observatory Satellite} study, the light curves of ST\,LMi and VV\,Pup showed an on/off, single-peaked structure over their \pspin, consistent with the accretion region being occulted by the WD as it spins. The short duration of the bright phase ($<0.5$\pspin\ for both sources) implies that the main accretion region is the lower pole \citep[see e.g.][]{Patterson-1984}. Conversely,  AM\,Her typically shows a much broader bright phase, indicative of predominantly upper pole accretion \citep[e.g.][]{Matt-2000}. However, the X-ray light curves of these `two visible (but one accreting) pole' polars have been seen to change substantially between visits by X-ray observatories, hinting at a more complex accretion geometry. AM\,Her, for example, has shown a so-called `reverse' soft X-ray mode, where the hard X-rays were modulated as normal, but the soft X-rays were anti-phased \citep{Heise-1985,Schwope-2020}, implying that the lower pole is also accreting, though perhaps not always. In addition AM\,Her has also shown a near disappearance of the soft X-ray modulation \citep{Matt-2000}, and occasionally a complete disappearance of any modulation in both soft and hard X-rays \citep{Priedhorsky-1987}.

\begin{figure}
    \centering
    \includegraphics{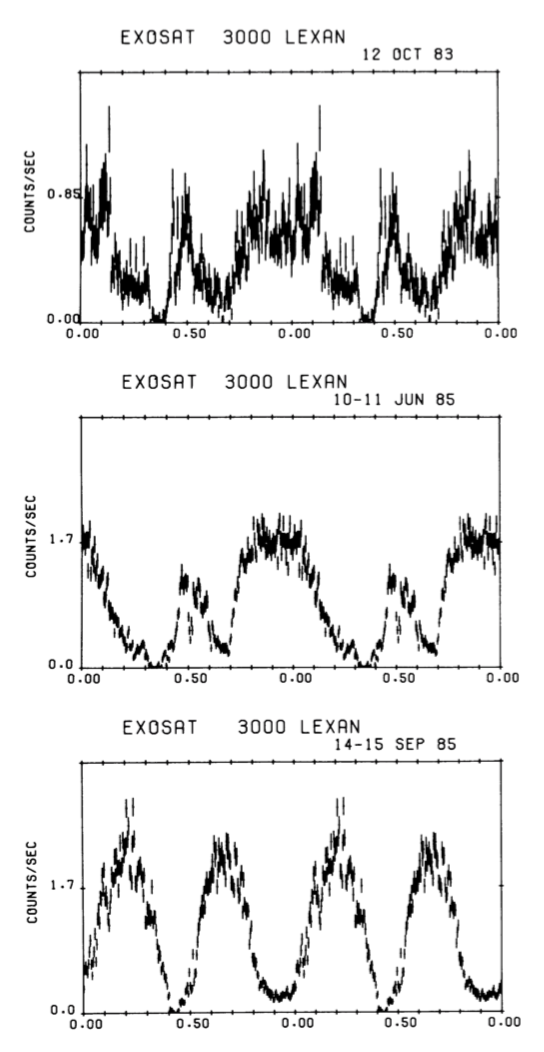}
    \caption{Soft X-ray (0.06--1.9 keV) light curve of the polar QQ\,Vul (E\,2003$+$225), as observed by {\em EXOSAT} at three different epochs from 1983--85. The top panel shows the complex, flaring behaviour exhibited by a number of polars, overlaid on a periodic modulation with two maxima. The middle panel is similar, though at a higher flux. In the top two panels, the X-ray flux drops to zero at phase $\phi\sim0.3-0.4$, interpreted by \cite{King-1985} as the magnetic stream occulting the soft X-rays from the upper pole. The bottom panel shows a completely different morphology, with two similar amplitude maxima at $\phi\sim0.25$ and $\phi\sim0.65$ and two uneven minima. 
    The changing morphology of the soft X-ray light curves of this and other polars hints at complex accretion geometries in these systems that are still not yet fully understood. Adapted with permission from \cite{Osborne-1987}.}
    \label{fig:QQVul}
\end{figure}

Other polars exhibit hard X-ray light curves that are always `on', though showing a sinusoidal modulation on \pspin, suggesting that either (a) these systems are accreting onto one pole (the upper pole), which is always in view, or (b) both poles are accreting and visible to the observer (\citep{Mason-1985} suggests the former). The soft X-ray light curves of these particular types of polars (EF\,Eri, AN\,UMa, QQ\,Vul and V834\,Cen are examples) are complicated and have been known to exhibit different properties across different epochs (see Fig. \ref{fig:QQVul}), similar to AM\,Her. However, the four sources studied by \cite{Mason-1985} all show a similar drop to zero flux at phase $\phi\sim0.3-0.4$ that lasts $\sim0.1$ cycle, which has been interpreted as the soft X-rays being occulted by the magnetic stream at the upper pole \citep{King-1985}. Similar behaviour is exhibited by the polar RX\,J1846.9$+$5538, which also appears to switch between phases of accreting onto one and both poles \citep{Schwarz-2002}. The complex soft X-ray light curves of these particular polars are hard to reconcile with the (once standard) single emission region models. \cite{Mukai-2017} suggests that a complex magnetic field geometry could feasibly lead to a light curve where each peak originates from a distinct emission region. Alternatively, \cite{Hameury-1988} consider instantaneous accretion of multiple soft X-ray emitting `blobs' to reproduce the strong flickering nature of the light curves.

One striking feature of the long-term light curves of polars is that they often undergo transitions to so-called `low states'. These states are characterized by significant ($\sim2$ orders of magnitude) drops in X-ray, optical and UV flux when compared to the average. \cite{Ramsay-2004b} examined \xmm\ observations of 37 polars and found 16 of them to be in a low state, with six of those not detected at all. The authors also reanalysed \rosat\ observations of polars and found that 57\% of that sample were also in a low state, implying that polars have a high/low state fractional duty cycle of $\sim50$\%. Though accretion activity is diminished in low-state polars, it is still occurring. Short-lived flares were seen at optical and X-ray wavelengths from UZ\,For during its 2001 low state, indicative of an accretion event \citep{Pandel-2002}. In addition, \cite{Ramsay-2004b} found that 8 of the 10 detected low states in the \xmm\ data showed significant variability associated with sporadic accretion on to the magnetic poles. Though the transitions between low and high states are likely to be related to changes in the mass-transfer rate in the system, it is unclear what causes the variations. \cite{Livio-1994} proposed that starspots on the secondary passing across the $L_1$ Langrangian point could be responsible for a decrease in mass-transfer rate. 
\cite{Wu-2008}, on the other hand, argue that variations in the magnetic field configuration in the WD could cause the observed pattern of alternating high/low states in AM\,Her. However, the true reason behind what causes the high and low states in polars is still uncertain.

\subsubsection{X-ray light curves of Intermediate Polars}
\label{sec:IP_LCs}

\begin{figure}
    \centering
\begin{overpic}[width=0.35\textwidth, angle=-90]{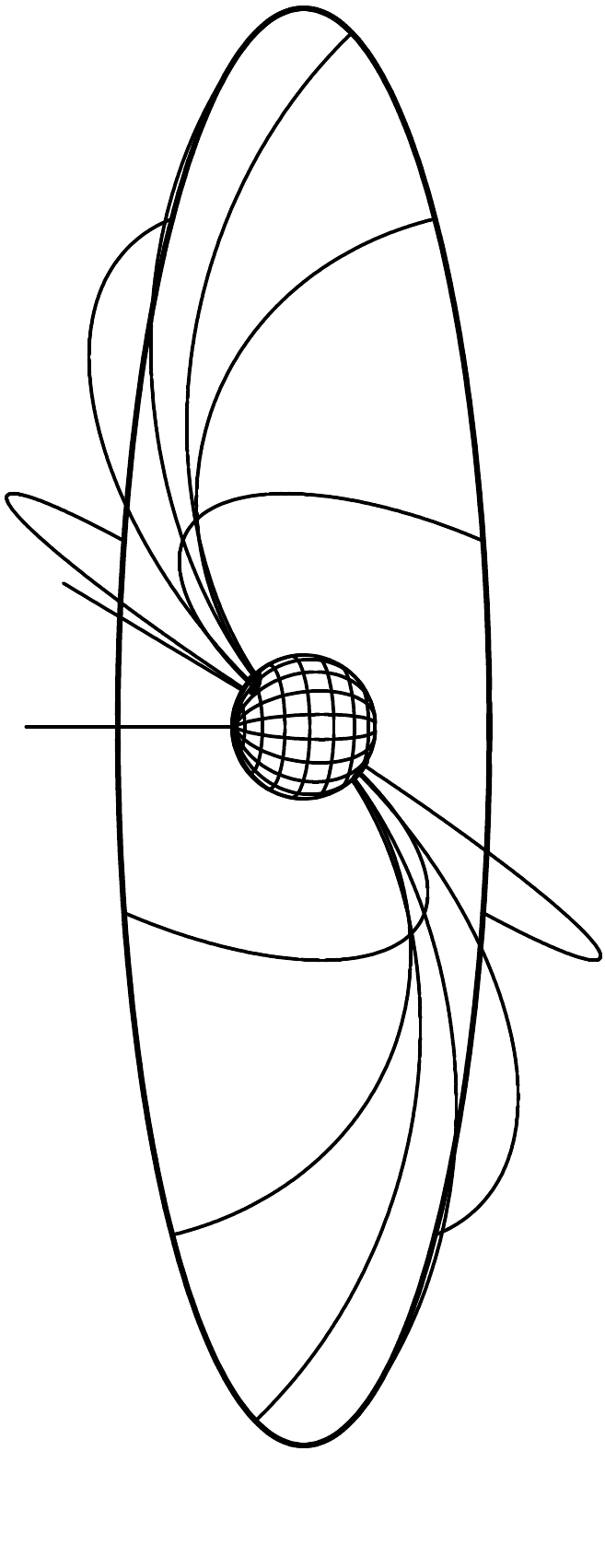}
\put(42, 35){\color{black}Spin axis}
\put(56, 38){\color{black}Mag.}
\put(56, 35){\color{black}dipole}
\put(56, 32){\color{black}axis}
\put(26, 20){\color{black}Field lines}
\put(59, 21){\color{black}Upper pole}
\put(51, 12){\color{black}Lower pole}
\put(-3,20){\color{black}Plane of}
\put(-3,17){\color{black}truncated}
\put(-3,14){\color{black}accretion}
\put(-3,11){\color{black}disc}
\end{overpic}
\caption{WD accreting onto both upper and lower poles. The vertical straight line marks the spin axis of the WD, while the straight angled line shows the magnetic dipole axis. The curves signify the magnetic field lines along which the material will flow. Assuming the observer is positioned at the top of the diagram, above the plane of the accretion disc (shown by the ellipse), then the `upper pole' is the one at the top right, and the `lower pole' at the bottom left.}
\label{fig:ip_curt}
\end{figure}
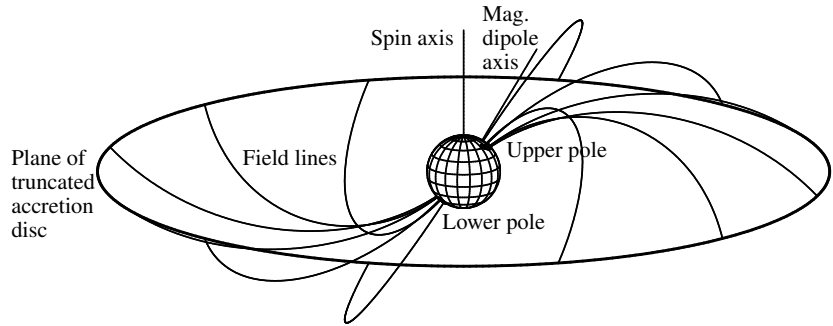

IPs, unlike polars, have central WDs that are spinning much faster than the \porb\ of the system. A typical IP will show a strong periodic modulation on the WD \pspin, as material is being forced onto field lines at a nearly uniform rate from a Keplerian disc, the inner edge of which is co-rotating with the WD. Similar to polars, spin-resolved light curves of IPs can also show self-occultation of the lower pole by the WD for part of the spin phase, depending on the viewing angle. Fig.~\ref{fig:ip_curt} shows a diagram of a WD in an IP system accreting onto both poles. 

The amplitude of the X-ray spin modulation in IPs is generally seen to decrease with increasing photon energy \citep{Norton-1989}. The original model to describe this phenomenon assumes that the accretion curtain acts as a photoelectric absorber, hence why the soft X-ray light curves show deeper minima \citep{Rosen-1988}. However, \cite{Norton-1989} showed that the energy dependence of the spin modulation of a number of IPs observed by {\it EXOSAT} was not as strong as expected from a single absorber. They find that any absorber with a high enough column density to produce a detectable oscillation in the 6--10 keV band would almost completely absorb all photons in the 0.5--2 keV band, yet this is not observed. However, if the simple absorber is instead replaced by an absorber that only partially covers the source of X-rays, then one can reproduce the observed modulation depths.

The true geometry of the absorber is complicated. \cite{Done-1998} found that the absorbing material (located in the pre-shock flow) requires a distribution of covering fraction as a function of column density. In the context of this model, the modulation seen in the X-ray light curves of most IPs can be explained as changes in the maximum column density of the complex absorber.

\begin{figure}[t!]
    \centering
    \includegraphics[width=\textwidth]{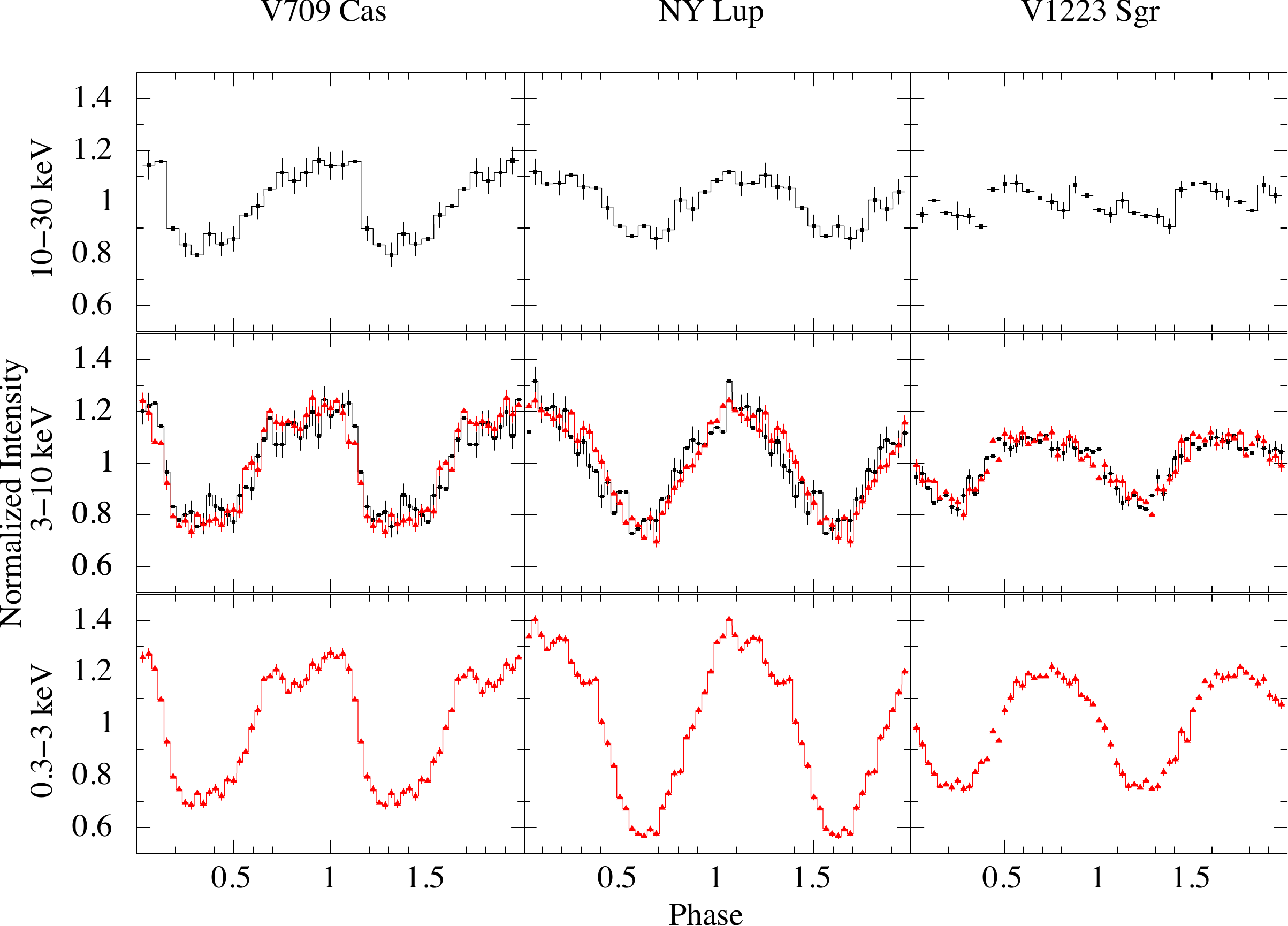}
    \caption{X-ray light curves of three IPs, folded on their \pspin, showing a decreasing amplitude as the photon energies increase. Data shown in red are from the EPIC-PN instrument onboard \xmm\ (0.3--10 keV energy range) and data in black are from \nustar\ (3--78 keV). Reproduced with permission from \cite{Mukai-2015}.}
    \label{fig:Mukai15_spin}
\end{figure}

However, the energy dependent absorption model predicts that there should be negligible absorption above 10 keV, yet some IPs have in fact shown modulations at these high energies. V709\,Cas and NY\,Lup both show modulations with a pulsed fraction $>10$\% in their 10--30 keV \nustar\ light curves \citep[see Fig. \ref{fig:Mukai15_spin};][]{Mukai-2015}. For these sources, though the complex partial covering model explains the modulations at lower energies, \cite{Mukai-2015} found that a non-negligible shock height could explain the observed $>10$ keV light curves. In this model, the slightly elevated shock means that, as the WD spins, there are a variety of viewing angles during which both X-ray emitting poles are simultaneously visible to the observer, hence the modulation \citep{Mukai-1999}. The finite shock height model can also be used to explain the properties of EX\,Hya, for which there is no evidence for absorption effects in the X-ray light curves \citep{Mukai-1999,Luna-2018}. \cite{Mukai-2015} showed that combining knowledge of the reflection properties of the X-ray spectra of IPs with the spin-resolved X-ray light curves can provide strong constraints on shock height.

Though the X-ray light curves for the vast majority of IPs show strong modulations on \pspin, the asynchronous nature of the systems means there can be complex interactions between the spin of the WD and the orbital motion of the binary. Thus, we often see modulations not just at \pspin\ and \porb, but also \pbeat~=~(\pspin$^{-1}$~$-$~\porb$^{-1})^{-1}$ and its harmonics. The relative strengths of the various modulations present in the power spectra of IP light curves (at both X-ray and optical wavelengths) are in fact an important diagnostic of the dominant mode of accretion in IPs \citep{Wynn-1992,Ferrario-1999}.

The classic picture of an IP, as discussed at the very beginning of Section \ref{sec:mag}, assumes that the WD is accreting from at least a partial accretion disc that terminates at \rmag. At \rmag, material then flows along the magnetic field lines on to the magnetic poles of the WD, hence we see a modulation on \pspin. An optical observer might also see modulations on \pbeat\ due to reprocessing of X-rays by parts of the system that are outside of the WD co-rotation frame (e.g. in the binary rest frame; \cite{Warner-1986}). However, what would we see if there were no partial accretion disc, and matter flowed directly from the donor on to the WD? The idea of a discless IP was first explored by \cite{Hameury-1986}, who drew comparisons to Algol systems. In many short-period Algols, the distance of closest approach of the ballistic gas stream that travels from $L_1$ ($R_{\rm min}$) is less than the radius of the accreting star, such that no disc can form (see also direct impact AM CVn systems in Section~\ref{sec:direct_impact}). Similarly, for some IPs, an orbital configuration might arise in which $R_{\rm min}~<~$\rmag, in which case no disc would form and matter instead flows along magnetic field lines in a similar fashion to polars, though still highly asynchronous. This is known as `stream-fed' accretion. \cite{Hameury-1986} estimated that the $R_{\rm min}~<~$\rmag\ inequality would be satisfied for IPs with \porb$<5$~h. A re-examination of the conditions required for discs in IPs concluded that if \pspin~$\lesssim0.1$~\porb\ then a partial disc can form \citep[][see also Fig. \ref{fig:spin-orb}]{King-1991}. If this inequality is broken, then either the system is severely out of spin equilibrium (theorized to be the case for EX\,Hya; \cite{Hellier-2014}), or the accretion flow is unlike the standard Keplerian disc we assume for IPs, hence the stream-fed model.\footnote{However, \cite{Norton-2004} find that, for IPs where \pspin/\porb$\gtrsim0.5$, accretion may instead be fed from a non-Keplerian ring-like structure at the edge of the WD's Roche lobe.} Furthermore, there is also a theorized hybrid between the stream- and disc-fed models known as the `disc-overflow' model in which a partial disc does form, but some of the ballistic stream is also able to overflow the disc at the same time and flow along the field lines \citep[e.g.][]{Hellier-1993}. For disc-overflow accretion to occur, the ballistic stream needs to approach the inner edge of the disc, which is the case when $R_{\rm min}\sim R_{\rm m}$. Thus, we might expect disc-overflow accretion to take place in systems where \pspin$\sim0.1$\porb\ (such as FO\,Aqr, see below).

\cite{Wynn-1992} developed a number of discless IP models and calculated theoretical X-ray power spectra. They found that for a simple stream-fed IP at low inclination (i.e. no eclipses), the majority of the X-rays are modulated on \pbeat\ rather than \pspin. This is due to the phenomenon of `pole-switching' where a given field line (fixed in the WD spin frame) will sweep up material from the accretion stream (fixed in the orbital frame), but the material is preferentially directed on to the nearest pole, which switches every half a beat cycle. \cite{Ferrario-1999} explored the concept further, finding that one can also use the optical light curves and spectra to distinguish between dominant accretion models. V2400\,Oph is widely regarded to be the first known discless IP, showing strong X-ray variations on \pbeat\ \citep{Buckley-1997}.

FO\,Aqr is the prototypical `disc-overflow' system -- essentially a hybrid accretion geometry -- as it often shows a strong X-ray modulation at both \pspin\ and \pbeat\ \citep{Hellier-1993}. However, the relative strength of the spin and beat components in the X-ray power spectrum has been known to change between observations, suggesting that the accretion geometry of the system, and perhaps other IPs, may be somewhat fluid. In 2016 FO\,Aqr showed a drastic drop in flux at both optical and X-ray energies \citep{Kennedy-2017}. The X-ray power spectra suggested that this so-called `low state' was linked to a transition to a stream-fed geometry. Upon returning to its typical flux, the WD spin began to dominate once more, suggesting a return to a typical disc-fed geometry. FO\,Aqr showed two more low states in 2017 and 2018, with each drop in flux associated with a change in the dominant accretion mode \citep{Littlefield-2020}, suggesting a fundamental link between mass-transfer rate and accretion geometry. Several IPs have historically shown low states that have been reported after the fact \citep[e.g.][]{Garnavich-1988,Shaw-2020b} so it has been difficult to study the X-ray timing properties of these systems. However, the decreases in flux in these sources may also be linked to a changing accretion geometry.

Though the light curves of IPs are typically dominated by the periodic variability discussed above, it is important to note that IPs do also show {\it aperiodic} variability. The aperiodic power spectra of IPs\footnote{The aperiodic power spectrum, or noise power spectrum, refers to the power spectrum of a light curve that has had known periodic variability removed} often show a broken power law shape, with a characteristic break frequency $\nu_{\rm b}$. The value of $\nu_{\rm b}$ is thought to be associated with the Keplerian frequency of the disc at its inner edge \citep{Revnivtsev-2009,Revnivtsev-2011}; i.e., for a typical disc-fed IP, at \rmag. If this is the case, then one can measure $\nu_{\rm b}$ to constrain \rmag:

\begin{equation}
    \nu_{\rm b} = \frac{1}{2\pi} \sqrt{\frac{GM_{\rm WD}}{R_{\rm m}^3}}
    \label{eq:break}
\end{equation}

Recalling, from Section \ref{sec:mCVspectra}, that Equation \ref{eq:freefall} can be modified by a $\sqrt{1-r_{\rm m}^{-1}}$ term for matter not falling from $\infty$, one finds that the value of \kts\ decreases with decreasing \rmag. If the inner disc is particularly close to the WD ($\lesssim4$\rwd; see e.g. \cite{Suleimanov-2019}), then failing to take this into account can result in an underestimate in \mwd. Thus, in some recent mass surveys of IPs, the aperiodic power spectra have been used to constrain \rmag\ and account for non-infinite fall heights in the subsequent calculations of \mwd\ \citep{Suleimanov-2016,Suleimanov-2019,Shaw-2020b}.

\subsubsection{AE\,Aqr and the propeller systems}
\label{sec:AEAqr}

For fast spinning and strongly magnetized mCVs, another interesting phenomenon can be observed. AE\,Aqr is an IP with $P_{\rm spin}=33$~s, and the WD is spinning down at a rate of $\dot{P}_{\rm spin}=5.64\times10^{-14}$~s~s$^{-1}$ \citep{deJager-1994}. A magnetized WD spinning at such a high rate should eject matter from the system as the rapidly rotating magnetosphere acts as a centrifugal barrier to any material flowing from the donor star \citep{Wynn-1997}. This `propeller effect' is something which is often seen in other accreting systems such as those containing neutron stars \citep[e.g.][]{Tsygankov-2016}, but for a long time AE\,Aqr was the only confirmed example of a CV in a propeller mode. Though the majority of matter is being flung away from the system by the magnetosphere, AE\,Aqr is still an X-ray source, and the X-ray emission is seen to be pulsed on the WD $P_{\rm spin}$, suggesting that some material is managing to penetrate the barrier and accrete onto the WD \citep[e.g.][]{Kitaguchi-2014}. 

There are other IPs that appear to show rapidly rotating WDs similar to AE\,Aqr, 
but these show no signatures of the propeller effect and are entirely consistent with being accretion powered. However, in 2020, LAMOST\,J024048.51$+$195226.9 was seen to exhibit similar flaring and spectral properties to AE\,Aqr, raising the possibility of it being an AE\,Aqr twin \citep{Garnavich-2021}. The initially elusive WD spin was detected at optical wavelengths with \pspin~=~24.93~s \citep{Pelisoli-2021} and firmly established LAMOST\,J024048.51$+$195226.9 as only the second IP seen in a propeller mode. Both AE\,Aqr and LAMOST\,J024048.51$+$195226.9 are also persistent radio sources \citep{Bookbinder-1987,Pretorius-2021}, which is rare for IPs \citep{Barrett-2020}. However, the origins of the radio emission are believed to be different for each source, with the radio emission in AE\,Aqr consistent with synchrotron emitting material, while the emission from LAMOST\,J024048.51$+$195226.9 is characteristic of magnetic plasma radiation or electron cyclotron maser emission, possibly from the magnetically active donor star \citep{Barrett-2022}.

Similar to AE\,Aqr and LAMOST\,J024048.51$+$195226.9, AR\,Sco is a WD binary with $P_{\rm spin}=1.95$ min and $P_{\rm orb}=3.56$ h \citep{Marsh-2016}. However, despite having very similar broadband and temporal properties to the two propeller systems, AR\,Sco cannot be considered an IP as there is no evidence for accretion in the system. Instead, AR\,Sco is thought to be a WD analogue to pulsars, with all non-stellar emission powered by the spin-down of the WD \citep{Buckley-2017}.

\section{AM\,CVn systems}

As discussed in the chapter by Webb, AM CVn stars are a subclass of CVs that consist of a WD accreting matter from a hydrogen-poor companion -- often another WD or a non-degenerate helium donor (see \citep{Solheim-2010} for a review). They have short orbital periods (\porb\ as low as 5 minutes in the case of HM\,Cnc; \cite{Israel-1999}). Historically, AM\,CVn systems have been relatively poorly studied in X-rays compared to other accreting WDs -- only three were detected in the \rosat\ All Sky Survey, with one being AM\,CVn itself \citep{Ulla-1995}. However, the launch of \xmm\ at the end of the 20th century prompted a number of multi-wavelength studies.

As with all CVs, AM\,CVn systems accrete via Roche Lobe Overflow and all but two systems (discussed below) form an accretion disc. As a result, the X-ray emission mechanisms are similar to `normal' (i.e. hydrogen-rich) CVs; that is, the disc is never hot enough to emit X-rays, but they are instead produced through shocks at the BL between the inner disc and the WD accretor.

The X-ray spectra of the AM\,CVn stars (with discs, at least) are well fit with a multi-temperature thermal plasma model \citep[see, e.g.,][]{Ramsay-2005}. However, unlike hydrogen-rich CVs, the elemental abundances deviate significantly from a solar composition. Some systems \citep[e.g.][]{Strohmayer-2004a,Ramsay-2005} show significant nitrogen and neon overabundance in addition to the expected hydrogen deficiency. A nitrogen overabundance can be explained by CNO processes in the donor \citep{Marsh-1991}. However, a neon overabundance can only be explained through the helium burning process, where nitrogen is burned into carbon and oxygen, so an excess of both nitrogen and neon is puzzling. \cite{Kupfer-2016} postulate that, in the case of GP\,Com, a short phase of He burning may have occurred in the donor, but stopped before N was depleted and C/O became too abundant. Alternate scenarios involve crystallization processes in the core of the donor that enhance neon abundances. 
However, ultimately the authors found no satisfactory solution to the observed high neon and nitrogen abundances in GP\,Com, so it remains an open question.

The X-ray light curves of AM\,CVn systems generally show no coherent modulations \citep[e.g.][]{Ramsay-2005}, typical of systems with low magnetic field strengths. However, \rosat\ observations of GP\,Com did reveal flux and hardness ratio modulations on the \porb\ of the system \citep[e.g.][]{vanTeeseling-1994} which may indicate variable absorption due to the rotating accretion stream. The temporal properties of GP\,Com in X-rays remain an outlier in terms of AM\,CVn systems with discs, however.

\subsection{HM\,Cnc and V407\,Vul: Direct impact accretion}
\label{sec:direct_impact}

HM\,Cnc and V407\,Vul are a special case in the context of the AM\,CVn systems. They are the two shortest period binaries, with $P_{\rm orb}$~=~321 and 569~s for HM\,Cnc and V407\, Vul, respectively. Their X-ray spectra are very similar, exhibiting a soft, relatively featureless, BB-like spectrum, with enhanced neon (relative to solar) in V407\,Vul \citep{Strohmayer-2008,Ramsay-2008}.

\begin{figure}[t!]
    \centering
    \includegraphics[width=\textwidth]{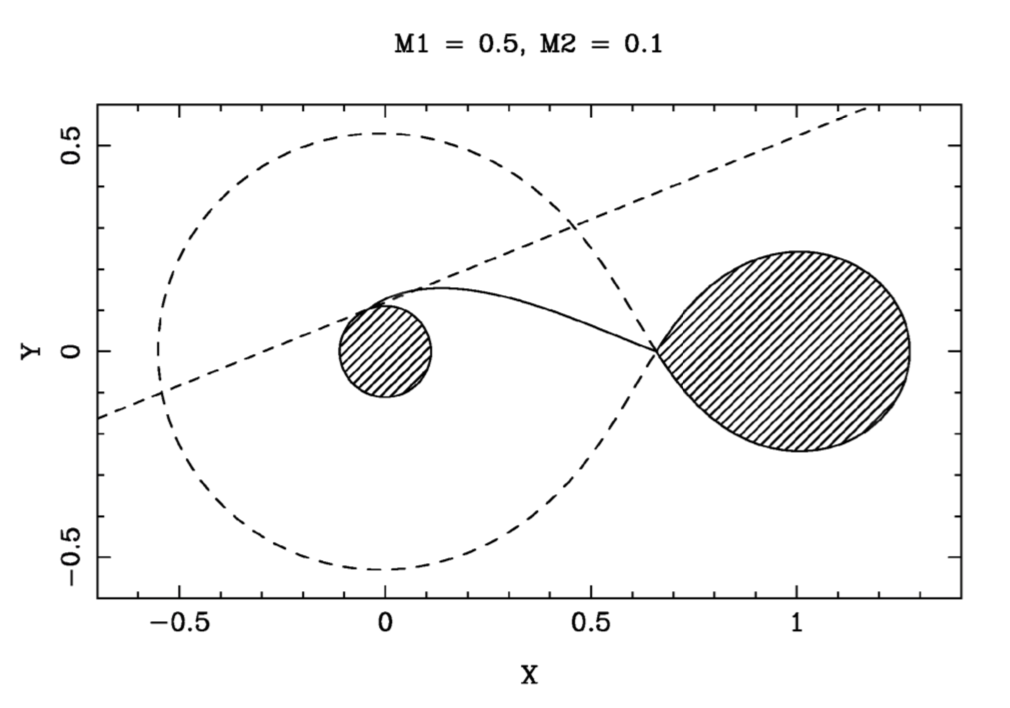}
    \caption{Schematic of the direct-impact model in the case of V407\,Vul, assuming a primary mass $M_1=0.5$ \ms\ and a donor mass $M_2=0.1$ \ms. The donor (center of mass at $X=1$) fills its Roche Lobe, and material flows through the $L_1$ point but impacts the surface of the primary before being able to form a disc. The dashed line is tangent to the impact point, showing that impact is hidden from the donor in this particular scenario. Reproduced with permission from \cite{Marsh-2002}}
    \label{fig:direct_impact}
\end{figure}

The short orbital periods of these two systems means that $R_{\rm min}$ (the distance of closest approach of the ballistic gas stream from the accretor; see Section \ref{sec:IP_LCs}) is smaller than the radius of the accretor itself, and the matter flowing from the $L_1$ point directly impacts the photosphere of the primary before it can form a disc \citep[see][and Fig. \ref{fig:direct_impact}]{Marsh-2002,Ramsay-2002b}. This `direct-impact' accretion model explains the majority of the observed properties of HM\,Cnc and V407\,Vul with the fewest assumptions.\footnote{ES\,Cet ($P_{\rm orb}=620$s) is sometimes classified as a direct-impact accretor (see e.g. discussion by \cite{Solheim-2010}) but the detection of double-peaked emission lines in the optical spectrum suggests that there is at least a small disc in the system \citep{Bakowska-2021}.} In the direct-impact model, the narrow accretion stream will impact an area smaller than a fraction $f=8.5\times10^{-5}$ of the surface area of the primary \citep{Dolence-2008}. The density and ram pressure of the stream is high enough to penetrate $\sim100$ km \citep{Wood-2009} into the photosphere of the accretor where it will thermalise, mix and well up to the surface to produce the observed soft X-rays (similar to the blob accretion model for soft X-ray emission in polars discussed in Section \ref{sec:soft_component}).

Unlike disc-fed AM\,CVn stars, the two direct impact systems do show strong coherent modulations \citep{Strohmayer-2002,Strohmayer-2005}. The light curves are characterized by an `on/off' pattern in the X-rays that lag the optical light curves by $\sim0.2$ phase. \cite{Wood-2009} showed that the X-ray light curves of HM\,Cnc and V407\,Vul can be reproduced if there are two X-ray hot spots on the surface of the accretor, one for the impacting material and a second downstream of the impact point for the upwelling material. The slight phase offset of the X-ray light curves with respect to the optical emission is thought to be due to a temperature gradient around the impact/upwelling regions, with temperature decreasing as a function of distance from the hotspots \citep{Marsh-2002,Wood-2009}.

\section{Conclusions}

Cataclysmic variables are binary star systems, each of which consists of a WD accreting from a companion star; yet, despite the similar constituents, they form a diverse group. As the name suggests, CVs vary, both in the optical and X-ray wavebands. Soft X-rays are emitted through nuclear burning on the surfaces of some WDs, while other systems show harder X-ray emission arising from shock interactions; some CVs show both lower- and higher-energy X-ray photons. The WD mass, accretion rate and magnetic field strength are defining characteristics of CV evolution, and lead to the range of observed properties. 

This chapter has provided an introduction to the X-ray emission from accreting WDs. This is a broad field, with an ever-increasing population, as X-ray missions become more sensitive, and further all-sky surveys are performed. The `Living {\em Swift} X-ray Point Source' (LSXPS; \cite{lsxps}) Catalogue is now expanding upon the previous static {\em Swift} X-ray catalogues, updating constantly with every new {\em Swift} observation, with a corresponding list of transients serendipitously (and automatically) discovered in the data -- some of which will be new CV outbursts. In addition, upcoming missions such as XRISM\footnote{X-Ray Imaging Spectroscopy Mission} and Athena\footnote{Advanced Telescope for High Energy Astrophysics} will offer high-resolution spectroscopy of CVs and symbiotic stars, opening up a new parameter space for research.

\bibliography{ref.bib}

\end{document}